%% file: main.tex
\documentclass[JEL]{AEA}

\usepackage{mathptmx}
\usepackage{natbib}
\usepackage{hyperref}       
\usepackage{url}            
\usepackage{amsmath}
\usepackage{booktabs}       
\usepackage{graphicx}
 \usepackage{xspace}
\usepackage{array}
\usepackage[neverdecrease]{paralist}

\newcommand{\ndjv}{\texttt{News D\'ej\`a Vu}\xspace}
\newcommand{\lt}{\texttt{LinkTransformer}\xspace}
\newcommand{\effocr}{\texttt{EffOCR}\xspace}

\newcommand{\lp}{\texttt{Layout Parser \xspace}}

\usepackage{xcolor, soul}
\sethlcolor{lightgray}

\title{Deep Learning for Economists}

\author{Melissa Dell\thanks{%
Department of Economics, Harvard University and NBER, melissadell@fas.harvard.edu. I would like to thank four anonymous referees, the editor, and Jake Carlson for their helpful comments and suggestions. Yiyang Chen provided excellent research assistance.}}

\begin{document}

\begin{abstract}
    Deep learning provides powerful methods to impute structured information from large-scale, unstructured text and image datasets. For example, economists might wish to detect the presence of economic activity in satellite images, or to measure the topics or entities mentioned in social media, the congressional record, or firm filings. This review introduces deep neural networks, covering methods such as classifiers, regression models, generative AI, and embedding models. Applications include classification, document digitization, record linkage, and methods for data exploration in massive scale text and image corpora. When suitable methods are used, deep learning models can be cheap to tune and can scale affordably to problems involving millions or billions of data points.. The review is accompanied by a regularly updated companion website, \href{https://econdl.github.io/}{EconDL}, with user-friendly demo notebooks, software resources, and a knowledge base that provides technical details and additional applications. 
\end{abstract}
\maketitle

\section{Introduction}

Deep neural networks have led to many recent scientific achievements – ranging from landing a rover on rugged Martian terrain to creating capable chatbots to transforming the diagnosis of disease. 
Deep neural networks typically map \textit{unstructured} data - such as text, document image scans, satellite and other imagery, videos, and audio - to a continuous vector space. In other words, they map complex and diverse types of data into a format that is easier to process and understand.
In the above examples, the resulting vectors are used to compute the instructions to steer the spacecraft, autoregressively predict what word comes next given a prompt, or identify whether an image contains a tumor. Analogously, an economist might use neural networks to detect the presence of informal vendors in street view images, or to measure the topics or people mentioned in firm filings or government documents.

At its core, deep learning is an approach for learning representations of data from empirical examples \citep{lecun2015deep}. These representations simplify high dimensional unstructured data into continuous vectors.
Deep neural networks learn representations at multiple layers of abstraction, combining non-linear neural network modules that each transform the representation in the previous layer of the neural network into a slightly more abstract representation, using learned weights (the term ``deep'' signifies these many layers of transformation). These weights are estimated by minimizing a loss function that compares model predictions for some task to ground truth examples. 

Why would one use a neural network to transform the raw data into these vector representations, versus just working directly with raw texts or images? First, deep neural networks don't just learn from the problem at hand. Rather, they incorporate relevant information in their parameters from exposure to massive-scale data. During pre-training, a modern language model, or vision model, will have been exposed to many millions of texts or images, learning the basic structures of language or vision. Exposure to vast amounts of data is essential to strong performance when processing unstructured data, because human language and vision are remarkably complex. This principle is called \textit{transfer learning} and is core to the success of deep neural methods.

\input{figures/flowchart}

Moreover, raw pixels or words lack \textit{context}, which is necessary for interpreting their meaning. Deep neural networks provide a powerful method for computing \textit{contextualized representations}. 
They map terms or pixels to vectors that depend on other nearby terms or pixels, with parameters learned mostly through massive-scale pre-training. 

Finally, raw texts and images are computationally unwieldy. In contrast, there are extremely optimized tools for continuous vector computations. For example, \citet{silcock332noise} make $10^{14}$ exact vector similarity calculations on a single mid-range GPU in 3 hours. This means that data can be analyzed at an \textit{unprecedented scale}. Theories are tested with data, and while more data won't solve challenges of causality, in general, it will provide economists with more fine-grained information for testing various hypotheses.

This review aims to bridge the gap between state-of-the-art deep learning research and economic applications. It focuses on imputing low-dimensional structured data from unstructured texts or images, in contexts where the ground truth is uncontroversial but extraction needs to be automated due to the massive scale of the problem. This structured data is then used for causal or descriptive analyses, whether as an outcome, endogenous variable, instrument, or control. Tasks that economists already perform by hand or with traditional methods (\textit{e.g.,} record linkage, text classification, document scan digitization) can be more accurately automated at scale, and deep learning also facilitates the extraction of novel data. Like a prior review by \citet{gentzkow2019text}, this review emphasizes text as data, but with methods developed since the publication of that article.

Many of the applications in this review fall under the broad umbrella of \textit{classification}: mapping high-dimensional unstructured data into discrete classes. The classes could identify types of objects present in a satellite image or numbers and words in a document image scan. Alternatively, classes could identify the topics of texts, the underlying source they were reproduced from, or unique individuals or locations referenced in them. A language model can be used to encode the raw text into lower-dimensional dense vector representations, one for the full text and one for each individual term (where ``dense'' means that the vector has a non-zero value in every position). The researcher can use these vector representations to predict whether the text is about a given topic, which locations are referenced, etc., by adding a classifier layer to the language model. Image classification works analogously. Generative AI models can also be prompted to impute these classifications. Alternatively, one can work directly with the dense vector representations, which are referred to as \textit{embeddings}.

Figure \ref{flowchart} provides a flow chart for approaching classification. The first question to ask is: ``Are the classes enumerated ex ante?" Sometimes, classes are not known, or the researcher may wish to add classes when applying the model to new settings without re-training it. A classifier computes a score for each class using the last layer of the language or vision model. Therefore, it can only be estimated when the classes are specified and seen in training. If classes are not specified, the researcher will need to work directly with the embeddings. If there are many classes, \textit{e.g.,} as with record linkage, where each unique entity can be thought of as a class, the researcher will again need to work with embeddings due to computational constraints in estimating a classifier.

When classes are specified ex ante and modest in number, either a classifier or generative AI may be well suited to the problem. If applications diverge from the data used to pre-train the neural network—common when working with historical data, document scans, or certain specialized settings—there is significant \textit{domain shift} from the pre-training corpus, and it may be necessary to tune a customized classifier to achieve strong performance. The nuance of the class definitions is also important. For straightforward tasks, framing classification as text generation using an off-the-shelf generative AI model like OpenAI's GPT may work well. For more nuanced tasks, a custom-trained classifier can better capture that nuance by being exposed to fine-grained examples. If in doubt, a researcher can try an off-the-shelf method and switch to a customized classifier if performance is unsatisfactory. This review shows that while custom-trained classifiers most often outperform GPT on text classification tasks, generative AI and custom classifiers both perform well on straightforward tasks. The review also considers the costs of these approaches.

\input{tables/applications}

Table \ref{tab:apps} summarizes this review's applications. Most can be framed as classification problems. The article also reviews regression, where a neural network is used to impute continuous values from text or images. It is important to note that the term ``regression'' is used somewhat differently in machine learning than it is in economics, which can generate confusion. 
In machine learning, ``regression'' refers to the prediction of a continuous number, whereas ``classification'' refers to the prediction of discrete outcomes. Regression encompasses any type of model used to predict continuous outcomes; in this review, we consider regression using deep neural networks. 

In the pre-deep learning era, problems in different domains were approached in very different ways, using rules heavily engineered to the specific features of a given language or a particular type of image, etc.; whereas deep learning has a remarkable capacity to generalize. Natural language processing (NLP), computer vision, and audio processing all use the same state-of-the-art neural network architecture, for instance. This generalizability is apparent in the diverse applications discussed in this review.

A variety of neural network applications are beyond the scope of this review. It does not cover how language models can be used more generally to enhance economists' productivity, as discussed by \citet{korinek2023generative}. It also does not cover machine learning methods outside of deep learning, such as those applied to structured data (which typically use shallower networks), as summarized in a review by \citet{athey2019machine}. Additionally, it does not examine using deep neural networks to compute approximate solutions to combinatorial optimization and high-dimensional DSGE problems. Approximating these solutions requires learning a neural network to map the original problem to a continuous vector space that preserves the essential properties of the problem. This is useful because computing an approximate solution in this space is dramatically faster than with traditional methods, allowing much larger problems to be approximated. There are many parallels with the methods covered in this review, but the applications are different enough that they necessitate their own treatment. Readers may wish to consult courses by \citet{fernandez2023deep} and \citet{vitercik2023machine}. Finally, a small literature directly uses deep neural networks in a causal framework. For instance, \citet{lynn2020causal} use classifiers and experiments to examine how variations in texts causally influence decision-making. While there is a role for this when experimentally manipulating text, often a researcher would like to extract low-dimensional representations from high-dimensional unstructured data (\textit{e.g.,} a text's topic, objects in a satellite image, numbers in a table scan, which textual records refer to the same firm) and use these—not the unstructured data—in the causal estimating equation. Hence, the focus here is on predicting these low-dimensional characteristics. The review does not attempt to summarize how these predictions have been used in economic studies, as this literature is new and rapidly evolving.

The reader may be wondering how quickly this review will become outdated. It is helpful to consider the popular metaphor that neural networks are like Legos: different neural network components can be configured in various ways to achieve different ends, or to achieve a more state-of-the-art version of the same end. This review focuses on frameworks where it is straightforward to swap in new neural network components as the literature advances, \textit{e.g.,} replacing an older convolutional neural network with a vision transformer \citep{dosovitskiy2020image}, or updating a BERT language model backbone \citep{devlin2018bert} with the most recent language model. Technical and implementation details—those most likely to change as the literature advances—are provided on the accompanying EconDL website: \url{https://econdl.github.io/}. It provides a knowledge base organized into core topics, as well as links to open-source packages geared towards economists and pipelines that construct large-scale datasets with deep learning. Interested readers will find lecture notes and links to blog posts, textbook treatments, open courseware, and original papers. EconDL also links to demo notebooks for many of the applications in this review. The website will be updated on a continual basis for as long as the transformer-based methods covered in this review remain state-of-the-art, and some of the packages explicitly support swapping in new neural networks as the literature advances.

This article is organized as follows: Section \ref{overview} provides an overview of deep learning, and Section \ref{architectures} introduces foundational architectures. 
Section \ref{Sec:data} discusses data requirements of deep learning, Section \ref{bias} considers bias and uncertainty quantification, and Section \ref{Sec:OpenDL} addresses reusability and reproducibility.  
Next we turn to applications. Section \ref{Sec:classification} introduces classification problems where the classes are defined ex ante and there are not too many classes, comparing classifiers and generative AI. 
Next, Section \ref{sec:metric} delves into embedding models, which are useful when the number of classes is large or the classes are not specified ex ante. 
Section \ref{Sec:regression} considers regression problems. There are other ways to approach the applications covered in this review. Section \ref{sec:OtherMethods} highlights why the methods emphasized are most likely to be suitable to the constraints faced by academic researchers. 
Section \ref{Sec:conclude} concludes.

\section{An Overview of Deep Learning} \label{overview}

\subsection{What is deep learning?}
Deep neural networks learn representations of raw data that extract information useful for specific tasks. Deep learning uses neural networks with many layers to map raw data to these representations, simplifying high-dimensional unstructured data into continuous vectors.

To represent data meaningfully for a given task, nodes (the numbers in a vector representation) in one layer of the neural network are transformed into nodes in the next by combining them with a non-linear function whose weights are learned parameters. These parameters—numbering millions to billions—are estimated by minimizing a cost function that compares model predictions on some task (\textit{e.g.,} predicting masked terms in text) to ground truth examples. For those unfamiliar with neural networks, I recommend the introductory videos by \citet{3blue1brown}.

The development of novel architectures and methods has made it feasible to optimize neural networks with millions to billions of parameters. These advances, while largely beyond the scope of this relatively broad review, are discussed in the EconDL knowledge base. In particular, many pioneering contributions in estimating deep networks were made in the literature on convolutional neural networks and are discussed in that post in the knowledge base.

Training a deep neural network from scratch requires a massive amount of data, and a couple of large-scale datasets are mainstays of the literature: ImageNet—a 14 million image dataset for image classification and related tasks \citep{deng2009imagenet}—and crawl corpora (\textit{e.g.,} Cleaned Colossal Common Crawl \citep{2019t5, dodge2021documenting})—massive public domain text datasets that essentially take a snapshot of the internet. Commercial models behind an API may also license proprietary training data. Training a deep neural model from scratch can require up to millions of dollars in compute, but fortunately, this is rarely necessary.

\input{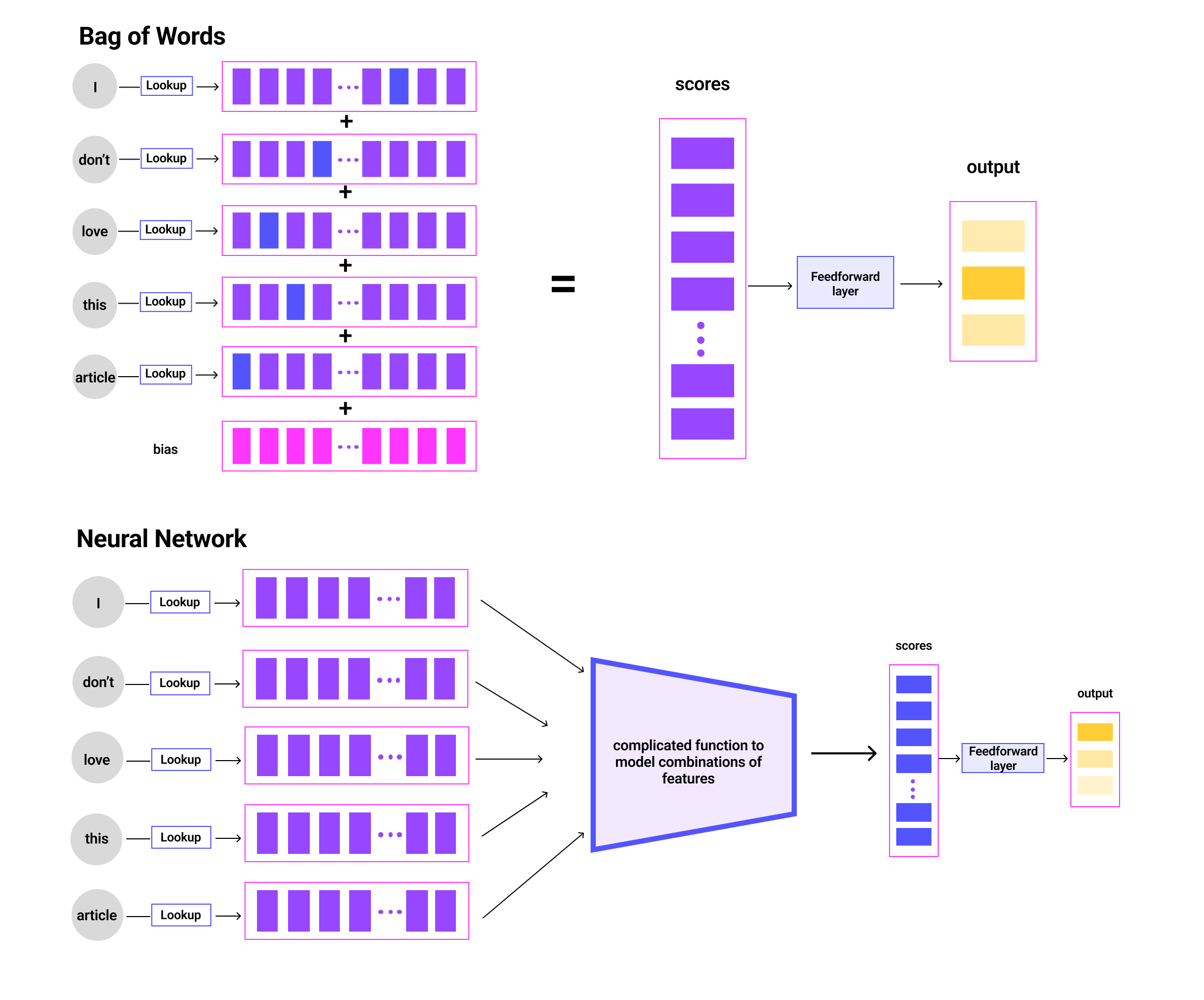}

Deep learning has transformed many fields because of the power of \textit{transfer learning}: deep networks trained in one domain can be adapted to many other domains with far fewer empirical training examples (often a few hundred to a few thousand) than would be required to train a model from scratch. For example, a researcher who needed to train a topic classifier could go to Hugging Face—the central hub for NLP (Section \ref{Sec:data})—and download a pre-trained state-of-the-art language model made publicly available by entities such as Google, Meta, and Microsoft. The language model was trained on a colossal corpus to predict tokens (words or sub-words) randomly masked from text. From this training, it learned to produce meaningful contextualized vector representations of texts. The researcher could add a classifier layer to the pre-trained language model and fine-tune the resulting neural network on their classification task using a relatively modest amount of labeled data. Most of the millions of model parameters would remain unchanged, as the model's basic understanding of language doesn't need updating, but the parameters that are of greatest relevance to the task at hand will update to improve model predictions \citep{merchant2020happens}.

One of the striking findings that has emerged from deep learning in recent years is that returns from increasing model size can continue to accrue even with very large models (e.g., billions of parameters) \citep{raffel2020exploring}. Human vision and language are complex, and a rich expressive model is needed to capture that complexity. 
For example, suppose we wish to perform the simple task of classifying which of ten objects (a horse, car, etc) is contained in an image. The inputs to a classifier are the RGB values of each pixel $x_{1,1}, x_{2,1}...x_{n,n}$.  
Suppose we estimate a linear classifier using these inputs. The score for each class $j=1...10$ is $\beta_{j, 1, 1} x_{1,1} + \beta_{j, 1, 2} x_{1,2} \cdots + \beta_{j, n, n} x_{n,n} + \gamma_j$, where the $\mathbf{\beta}_j$ and $\gamma_j$ are estimated parameters. The classifier predicts assignment to the class with the highest score.
The weight parameters for a class will be larger for pixels where the object in that class tends to be located. 
This is inherently brittle, because for each class there is only one parameter per pixel, but the horse could vary in its pose, its size, its position in the image, its color or build, etc.
In the plot of the $\mathbf{\beta_{horse}}$, one may see a horse standing in the middle of the image with two heads facing in either direction, as the linear classifier struggles to assign high values to pixels where horses are empirically likely to be located. 
A large neural network effectively allows for many such ``filters" in predicting whether a horse is in the image.

Alternatively, suppose we would like to analyze whether statements (\textit{e.g.,} in survey data) have a positive, negative, or neutral sentiment. A traditional approach, commonly used in the economics literature, is the bag-of-words method. The researcher looks up the sentiment of each word in a lookup table and aggregates these together to measure the sentiment of the sentence (Figure \ref{fig:bow}).

It is straightforward to see the limitations of this approach. Consider the following sentences: 'I love this article,' 'I don't love this article,' 'I don't hate this article,' 'There's nothing that I don't love about this article.' One cannot capture these different sentiments—even in these very simple sentences—by adding up independent representations of each word. Instead, we need to model nonlinear combinations of words, and neural networks are the state-of-the-art tool for approximating complicated nonlinear functions.

When processing text with a modern language model (bottom panel of Figure \ref{fig:bow}), a tokenizer first maps each word in the input to a number assigned by a lookup dictionary (if the word isn't in the dictionary, it gets split into subwords that are). These numbers get converted into vectors using learned parameters and are then passed through the neural network, which transforms them incrementally at each layer into semantically rich representations of the input tokens. Vision models are broadly analogous, taking pixels or patches of images as inputs. 

The main alternative to neural networks is to use human-engineered features. In other words, the researcher pre-specifies rules for processing raw information. For example, table digitization could be automated by writing rules to detect the connected white space that separates rows and columns. With deep learning, the model is instead shown annotated examples of table layouts. The deep learning revolution has illustrated over and over—across many different tasks—that learning from empirical examples greatly outperforms human-engineered feature extraction in processing unstructured data. Some of this evidence is discussed in the EconDL knowledge base.

Deep learning is likely to outperform feature engineering in many economic applications as well. The information that economists would like to process is frequently complex and noisy. For example, noise is introduced into document scans through aging, scanning, and historical printing techniques; alternatively, text data may contain OCR errors or typos. Human language is complex, with many different ways to express the same sentiment and words that can change meaning significantly depending on the context. Noise and complexity create exceptions to human-engineered rules, which must also be hard-coded, and likewise there are exceptions to the exceptions. What initially seems like a simple task can quickly become convoluted as the researcher tries to hard-code these exceptions. Even if the results are satisfactory, the human-engineered system is likely to be heavily tailored to the case at hand and will not translate well to other data with different types of complexity and noise.

Another potential advantage of deep learning is that recipes for training and implementing neural networks are standard and reproducible, whereas significant discretion is inherent in human-engineered feature extraction. Even leaving aside researcher degrees of freedom, significant domain knowledge is required to engineer rules. For example, in statistical machine translation, large numbers of researchers worked over decades to engineer complicated statistical rules for machine translation. These systems were outperformed by a neural network that a few researchers developed over a few months. Subsequent advances in neural translation led to the Transformer architecture \citep{vaswani2017attention}, which has since revolutionized natural language processing, computer vision, audio, and other domains.

\section{Foundational Deep Learning Architectures} \label{architectures}

This section provides a brief introduction to neural network architectures. For readers who are not familiar, I recommend consulting the EconDL knowledge base and the resources available there for a more detailed treatment.

\subsection{The Basics of Neural Networks} \label{arch:vanilla}

A neural network consists of layers of interconnected nodes, which are called \textit{neurons}. Each neuron holds a value. The value of a neuron is computed by combining values from neurons in the previous layer using an activation function and learned weights. These many layers transform the input (\textit{e.g.,} tokenized text) into vectors that are useful for performing a desired task.

An example activation function is rectified linear unit (ReLU): $f(x)=max(0, x)$, where 

\begin{equation}
x = w_1 \cdot i_1 + w_2 \cdot i_2 + \ldots + w_n \cdot i_n + b
\end{equation}

\( w_1, w_2, \ldots, w_n \) and \( b \) are the learned weights and bias terms and \( i_1, i_2, \ldots, i_n \) are the input values to the neuron from the previous layer in the network. When we feed data into a neural network, the input values are transformed by the activation functions at each layer. Nodes in the final layer are the output. 

Activation functions are an important component of neural networks, because they introduce non-linearity, enabling the network to capture non-linear relationships in the data. The Convolutional Neural Networks post in the knowledge base provides further introduction to activation functions. 

To optimize the neural network, the output is compared to ground truth labels using a loss function. What these labels measure depends on the objective: \textit{e.g.,} predicting masked terms for a language model or predicting image classes for an image model. 
As with any optimization problem, we need to know the gradient of the loss with respect to each weight to minimize the function. 
For each layer starting from the output layer and moving backward through the network to the input layer, we compute the gradient of the loss with respect to the weights in that layer. This requires using the chain rule.  The chain rule allows us to compute the derivative of the loss with respect to any weight in the network by multiplying together derivatives computed layer by layer. This is known as \textit{backpropagation.} Weights are adjusted using a gradient descent algorithm. 

Readers who are not familiar with backpropagation are encouraged to consult \citet{3blue1brown} for a high-production-value, graphical introduction. Readers wishing to gain a deeper understanding may enjoy \citet{karpathy_2022}, an advanced backpropagation tutorial. \citet{nielsen2015neural} provides a textbook treatment for those with no prior familiarity with neural networks. \citet{goodfellow2016deep} offers a textbook treatment for those already familiar who would like an in-depth review, and \citet{stevens2020deep} is aimed at those who would like to learn key concepts through hands-on implementations in PyTorch.

In a vanilla feedforward neural network, all neurons in one layer connect to all neurons in the next layer. Deep fully connected networks are rarely used in practice. Rather a few types of neural networks have dominated the deep learning literature. This review focuses on convolutional neural networks (CNNs; Section \ref{arch:cnn}), recurrent neural networks (RNNs; Section \ref{arch:rnn}), and the transformer (Sections \ref{arch:transformer}, \ref{arch:transformerllm}, and \ref{arch:vision}).

\subsection{Convolutional Neural Networks} \label{arch:cnn}

CNNs leverage the spatial structure present in images and played a central role in ushering in the deep learning revolution. Despite the advent of a newer architecture for image processing—the vision transformer (Section \ref{arch:vision})—they remain widely used, can obtain near-state-of-the-art performance when appropriately modernized, and can be lighter weight and easier to tune than vision transformers. This section provides a brief introduction. I recommend that those unfamiliar with CNNs consult the short graphical introduction to convolution by \citet{conv_mit}, as a visual introduction to the concepts described below is particularly helpful. Additional resources are on the `Convolutional Neural Network' page of the EconDL knowledge base.

Vision problems start with an image of a given height (in pixels), width (in pixels), and depth (\textit{e.g.,} 3 for an RGB image). Convolutional layers are the core building block of a CNN. The layer's parameters consist of a set of learnable filters, \textit{e.g.}, 3 $\times$ 3, 5 $\times$ 5, or 7 $\times$ 7 weight matrices. These filters are only applied to the nodes immediately surrounding a given node when computing the output for the next layer and extend through the full depth of the input. Each filter is convolved (moved) across the input, producing an activation at each spatial location. Using the same weights for different spatial locations drastically reduces the number of parameters compared to fully connected layers. 
Moreover, parameter sharing ensures that features can be detected regardless of their position in the image. This gives CNNs a degree of \textit{translation invariance}, desirable since \textit{e.g.,} a horse is a horse regardless of its position in the image.

The locality bias inherent in small convolutional filters is logical because the interpretation of a pixel is more influenced by its neighboring pixels than by distant ones. Despite the localized nature of these filters, a CNN still achieves an extensive receptive field through the depth of the network. CNNs are adept at learning hierarchical features: lower layers, which have a more limited receptive field, capture simple patterns such as edges, while deeper layers capture increasingly complex structures.

In addition to convolutional layers, CNNs also use pooling layers.
If $M$ different convolutional filters are applied to a neural network layer, the depth of the next layer will be $M$, since each filter produces an activation for each spatial location. Pooling layers reduce this depth, preventing the number of parameters from becoming infeasibly large. Often, a CNN consists of alternating convolutional and pooling layers. 

The central challenge to estimating neural networks with many layers is the vanishing gradient problem \citep{bengio1994learning}. Backpropagation computes the gradient of the cost function with respect to each weight in the network. This requires applying the chain rule to find the gradient of the loss with respect to the output of each layer, and then the gradient of the output of each layer with respect to its input. Derivatives can become very small for extreme values of their inputs. Backpropagation multiplies small gradients together. Hence, the gradient may become exponentially smaller as it flows back to the earlier layers. If the gradient becomes extremely small for the initial layers, learning will be very slow or stop altogether. The post on `Convolutional Neural Networks' on EconDL examines the evolution of CNN architectures, including key innovations that allowed for the optimization of much deeper, more expressive networks, circumventing the vanishing gradient problem \citep{krizhevsky2012imagenet, simonyan2014very, szegedy2015going, he2016deep, xie2017aggregated, howard2019searching, liu2022convnet}.

\subsection{Recurrent Neural Networks} \label{arch:rnn}

CNNs require fixed-size inputs, as neural networks are initialized with weight matrices of fixed dimensions (variably-sized images need to be resized or padded). In contrast, \textit{recurrent neural networks} (RNNs) are designed to process variably-sized inputs and outputs. They historically played an important role in NLP \citep{hochreiter1997long, greff2016lstm}, though they have since been superseded by the transformer. While researchers generally should use a transformer for NLP applications, I introduce RNNs as a point of comparison to the transformer.

RNNs process a sequence of inputs—\textit{e.g.,} tokens in a text—iteratively. At each time step, they maintain a state that captures historical information about the input sequence. This state is updated iteratively as the network processes each element in the sequence, allowing the network to `remember' previous elements in the variable-length input.

Long-range dependencies are important to human language. To use a prominent example from \citet{vaswani2017attention}, `The animal didn't cross the road because it was too tired' versus `The animal didn't cross the road because it was too wide.' Does `it' refer to the animal or the road? This depends on dependencies between `it' and other tokens in the input. The most prominent RNN is the bi-directional LSTM (Long Short-Term Memory) \citep{hochreiter1997long}. Bi-directionality captures dependencies in both directions by feeding the input sequence forwards and backwards. Readers can find a more detailed introduction to LSTMs in the EconDL knowledge base.

\subsection{The Transformer} \label{arch:transformer}

The \textit{transformer} \citep{vaswani2017attention} has revolutionized NLP and made inroads in nearly all areas of deep learning, including vision, audio, graphs, and reinforcement learning. For readers who are not familiar, I recommend the Illustrated Transformer blog post by \citep{alammar_transformer}, widely recognized as the most accessible introduction. An annotation of the original paper by \citet{rush2018annotated} is also a classic reference.

The original transformer was a neural translation model whose key ingredient is \textit{attention}. All tokens (words or sub-words) in a sequence are fed into the model in parallel, and the model attends to all other tokens in the context to create contextualized representations for each token. Contextualized representations contrast with traditional static representations of words \citep{mikolov2013distributed, pennington2014glove, olah2014deep}, where a given word in a training corpus always has the same representation. The transformer solves the locality bias of RNNs—where information is lost as the hidden state is passed along to each sequential token—because it can attend to any token in the sequence, whether nearby or not.

Self-attention is quadratic, which limits the length of text that can be passed into the model at one time  (the context window). A typical context window length in open-source models is 512 tokens. For many problems, this is sufficient (\textit{e.g.,} the texts can be chunked, or the first 512 tokens are sufficient to form a meaningful document representation). There are also models with sparse attention mechanisms that allow for long context windows.

Inputs are fed into the transformer in parallel, rather than sequentially as in an RNN, allowing training to fully leverage the parallel computing power of GPUs. This makes it computationally feasible to train much larger models on more data for longer, all of which improve performance \citep{raffel2020exploring}.

\subsection{Transformer Large Language Models} \label{arch:transformerllm}

Most modern NLP applications use transformer-based large language models (LLMs). For those unfamiliar with LLMs or needing a refresher, I highly recommend Jay Alammar's `Illustrated GPT-2/3' \citep{alammar_gpt2, alammar_gpt3} and `Illustrated BERT' \citep{alammar_bert} blog posts, which provide an intuitive graphical introduction to transformer LLM architectures.

There are two main types of transformer LLMs. \textit{Generative (decoder) models} predict the next word in a sequence \citep{radford2019language, brown2020language}. They are typically used for text generation. Since they are trained by predicting the next word, they can only attend to prior tokens when creating contextualized representations for a given token. This is called \textit{causal attention}.

\input{figures/BERT}

In contrast, \textit{masked (encoder) language models} are bidirectional: in creating contextualized representations of words in a sequence, they can attend to all words in the sequence (\textit{masked attention}). The model is trained by predicting masked tokens \citep{devlin2018bert, liu2019roberta, sanh2019distilbert,lan2019albert, he2020deberta}. Encoder models are typically used when a researcher aims to create representations of text that entail feeding an entire text to a model. Bidirectionality is helpful for such tasks, because the context both before and after a word is useful for creating semantically meaningful representations of it. Language models can also combine encoder and decoder transformer blocks, \textit{e.g.,} \citet{2019t5}.

With the transformer architecture, the same pre-trained language model can be used as the ``backbone'' for a wide variety of tasks, facilitating transfer learning. This is illustrated in Figure \ref{fig:bert}, which is adapted from an illustration in the original BERT paper \citep{devlin2018bert}. 
A transformer language model produces a vector representation for each token (word or sub-word) in its input, as well as a \texttt{<cls>} representation that represents the entire input text.
The first panel shows that full text sequences can be classified by adding a classifier head to the \texttt{<cls>} token. The classifier is a feedforward neural network that aggregates the nodes in the \texttt{<cls>} vector of the final transformer layer into a score for each class, using learned weights.
Alternatively, two texts can be embedded jointly - separated by the special token \texttt{<sep>} - and then a classifier can be added to the \texttt{<cls>} token to classify the relationship between texts (panel b). Or, each individual token can be classified (\textit{e.g.,}tagging whether it refers to an individuals, location, etc.) by adding classifier heads to each of the token embeddings (panel c). Spans of text can also be identified (\textit{e.g.,} the answer to a question; panel d). 

A variety of different transformer pre-trained language  models are detailed in the EconDL knowledge base post on `Transformer Language Models.'

\subsection{Vision and Audio Transformers} \label{arch:vision}
The transformer has transformed many other areas of deep learning, including computer vision \citep{dosovitskiy2020image, touvron2021training, grill2020bootstrap, caron2021emerging, ali2021xcit, he2022masked, chen2021empirical}. Vision Transformers (ViTs) use the same transformer architecture as transformer language models, with some adaptations to make them suitable for images. Unlike in NLP, where transformer language models have greatly outpaced the prior technology, in vision the gains from the transformer relative to CNNs are more modest \citep{liu2022convnet}. An appropriately modernized CNN will often be competitive with a similarly sized ViT. Moreover, the smallest CNNs (\textit{e.g.,} \cite{howard2019searching}) are smaller than lightweight ViTs (\textit{e.g.,} \cite{mehta2021mobilevit}) at present and can also perform well on straightforward tasks. Practically, I recommend starting with a lightweight CNN—which will be easier and significantly cheaper to train and deploy—and examining the performance of a larger CNN or ViT model if the lightweight model is inadequate. The EconDL post on Vision Transformers provides more details about ViT architectures.

The extent to which the transformer can be used to process highly diverse types of unstructured data is strikingly illustrated by its application to audio. State-of-the-art performance was achieved by applying a ViT to the spectrogram image of the audio \citep{gong2021ast}. Even more strikingly, performance was maximized by pre-training on ImageNet—the main benchmark for vision that consists of over 14 million natural images (\textit{e.g.,} of dogs, food, etc.)—a powerful illustration of transfer learning even across modalities.

\subsection{Optimizing Neural Networks} \label{arch:opt}

Being able to optimize a neural network is clearly central to using them in research. The optimizer (see \textit{e.g.,} \citet{kingma2014adam, goh2017momentum}), initialization \citep{he2015delving}, and normalization \citep{ioffe2015batch, santurkar2018does} are all important. To estimate deep neural models, the researcher must also select various hyperparameters (\textit{e.g.,} \cite{li2017hyperband, falkner2018bohb}). Interested readers are referred to the post on `Basics of Training and Optimizing Neural Networks' in the EconDL companion knowledge base. The packages on EconDL choose reasonable defaults for various hyperparameters in an effort to make training neural networks more user-friendly.

While there are various details at play, my main practical takeaway from guiding many students in optimizing neural networks is that when performance is unexpectedly poor, it is most often due to either a poorly chosen learning rate or incorrectly formatted input data. The learning rate determines the size of the steps taken by the optimizer while adjusting weights during training. If it is too low, the model won't update; if it is too high, the model weights will oscillate wildly. Moreover, data are expected in a specific format, and transformations may be performed on the fly (for instance, many neural networks require fixed-size inputs). Neural networks learn from empirical examples, and unexpectedly poor performance is often the result of feeding them misformatted examples. 

I also recommend that those new to deep learning train and deploy models using a cloud server specifically designed for this purpose. The EconDL tutorials use Google Colab. Installing deep learning packages locally requires resolving extensive dependencies, which may be challenging for those with limited experience. For more experienced users who heavily utilize deep learning in their research, purchasing their own GPUs can often provide significant cost savings.

\section{Training Data} \label{Sec:data}

High-quality training and evaluation data are integral to the utility of deep learning. In \textit{supervised learning}, data are partitioned into \textit{labeled} and \textit{unlabeled} sets. The unlabeled set, which consists of all the data that the deep learning model will be applied to, is typically much larger than the labeled set. Deep learning is \textit{self-supervised} when relevant labels are gleaned automatically from the data itself. For example, language models can be pre-trained by predicting words that have been randomly masked from a massive text corpus. Analogously, a masking strategy can be applied to images for the self-supervised pre-training of vision models \citep{he2022masked}. Self-supervised learning is most commonly used in pre-training, and then the pre-trained model is transferred to another domain and applied to unlabeled data (potentially following additional supervised tuning on a modest amount of data from the target domain). Finally, in \textit{unsupervised learning}, there are no ground truth labels, as the goal is to discover underlying structures in the data, grouping them according to similarities. Embeddings can be clustered, for example, to discover these relationships.

Supervised methods are common in economic applications, as the goal is often to use neural networks to extract some characteristics from unlabeled data.
Economists might also continue self-supervised pre-training. This is most common when the domain of their application shifts considerably from the domain that the model was pre-trained on \cite{gururangan2020dont}. For instance, an economic historian analyzing 18th century legal texts might first continue pre-training the language model on that database (by predicting masked tokens), in order to impart better understanding of 18th century legal jargon.

Unsupervised learning is most useful for data exploration. In empirical economics, the norm is to specify a narrowly defined hypothesis and test it statistically, which often lends itself to supervised applications. However, methodical data exploration using unsupervised methods can be a powerful tool for gleaning stylized facts from novel unstructured data.

When conducting supervised learning, labeled data are further divided into \textit{training data}—used to train the model, \textit{validation data}—used to tune model hyperparameters or select prompts, and \textit{test data}—used only to compute the model evaluations that will be reported in the findings. The researcher should always have a high-quality, representative test set to evaluate model performance. If the data used to evaluate model performance are not representative of the unlabeled dataset—and in particular, if some appreciable portion of the unlabeled data has no support in the labeled data—model performance on the evaluation data may diverge widely from model performance on the unlabeled data, which is the underlying object of interest.

In an ideal scenario, representative test sets can be created through random sampling. However, this is not always possible, particularly when classes that the researcher would like to measure are highly imbalanced. Suppose that a researcher needs to extract texts on a topic of interest from a massive web corpus, and the relevant topic appears only once in every ten thousand texts. The labeling requirements for sampling enough positives randomly are clearly infeasible. This scenario is common in social science, where researchers frequently need to classify relevant information from a massive corpus—\textit{e.g.,} media or government documents—where only a tiny share of content is about the topic of interest. 

While strategies to draw the most informative samples to annotate have generated a large machine learning literature on \textit{active learning} (EconDL provides a detailed discussion in the context of text classification), there is little work on selecting representative samples for training, evaluation, or debiasing when class imbalance is severe. Discriminative active learning \citep{gissin2019discriminative} selects samples to label that maximize the difficulty of distinguishing between the labeled and unlabeled data and can work well with relatively balanced data. It does not work well with severe class imbalance because it fails to sufficiently sample the rare class(es). Other active learning approaches seek to sample near the decision boundary of a classifier, which will sample the rare class(es) and can maximize predictive accuracy. However, this will provide an unrepresentative sample.

Social scientists, instead, frequently use the presence of certain keyword(s) to choose content to label. However, by construction, this fails to place positive sampling probability on all instances, increasing the odds that some types of unlabeled data have no support in the labeled data. This can generate prediction bias that is systematically correlated with the error term in the downstream causal estimating equation—where the researcher intends to use the deep learning model predictions—since semantics and omitted variables often both vary across space and time.

Embedding models (Section \ref{sec:metric})—deep neural networks that create a space where distances between vector representations of texts or images are meaningful—provide a metric that can be used for stratified sampling of data to label for training or evaluation. The closer a text/image is to a set of queries about a class (\textit{e.g.,} `this article is about tax policy'), the higher the probability that it comes from that class, making distances in this space useful for stratified sampling. A stratified sampling approach can also provide informative negatives for training: samples that a pre-trained model place near a query but that are not related to that query in the way the researcher intends. This is an active area of research, where economists have the potential to make important contributions. The EconDL site will update on this literature as it advances.

Training data need not be drawn from the same distribution as the unlabeled data—given the power of transfer learning—although predictive accuracy will typically decline with the magnitude of the \textit{domain shift} between the target and training data. Oftentimes, datasets that already exist or can be extracted from web texts allow for the cheap creation of a much larger training set than the researcher could label by hand. High performance on a target dataset is then ensured by further tuning on a much smaller set of hand-crafted labels from the target data.

\textit{Congruence labeling}—when two (or more) annotators label the same data points—is important for ensuring the quality of training and evaluation data. Even seemingly simple tasks are often messier than expected once taken to real-world unstructured data. Congruence labeling also ensures that annotators have understood the task and are producing high-quality labels. In challenging labeling tasks, researchers may wish for all labeled data to be double-annotated, resolving discrepancies by hand. In more straightforward cases, congruence labeling may only be necessary for a smaller subset of the data, to ensure that the task is well-defined and annotators have properly understood the instructions. In machine learning papers, the researcher is typically expected to report the congruence between annotators, as well as to publish annotator instructions, and this can be useful in economic applications as well.

\section{Bias and Uncertainty Quantification} \label{bias}

There are many limitations to using deep learning to solve social or economic problems (see, for instance, papers from the ACM Conference on Fairness, Accountability, and Transparency \url{https://facctconference.org/} and \citet{cui2022stable}). Here, our focus is much narrower: to impute or explore low dimensional features of unstructured data that humans are likely to agree on, in contexts where the size of the raw dataset is orders of magnitude too large to extract features manually.

Caution is necessary when applying deep learning to contexts that require subjective judgment. For example, researchers have shown that the self-identified political orientations of annotators influences their incongruence on political sentiment labeling \citep{shen2021what}. Sentiment classification in the deep learning literature has been designed largely around laptop, restaurant, and movie reviews, contexts where there is typically an explicit sentiment about the product that can be validated by stars. In many applications that economists care about—such as sentiment in media data, political speeches, corporate reports, etc.—sentiment can be much more implicit, and humans may not agree on it. If a model is fed annotations that reflect the subjective biases of the annotator—versus a well-defined ground truth—or if it simply does not have enough examples because the distinctions to be made are complex, it will make inaccurate predictions that may be systematically biased. Models can also inherit biases from pre-training, and there are large literatures on bias and fairness in AI \citep{mehrabi2021survey}. These challenges can be mitigated by sticking to straightforward tasks with a clearly defined ground truth.

Economists can make valuable contributions on uncertainty quantification, which is uncommon in much of the deep learning literature. Conformal inference can provide uncertainty quantification for prediction tasks. Facilitated by the collection of a ground truth calibration dataset, conformal methods produce prediction sets with marginal coverage guarantees under mild conditions. A canonical tutorial is \citet{shafer2008tutorial}; see \citet{chernozhukov2021exact}, \citet{cattaneo2022uncertainty}, and \citet{lei2020conformal} for recent contributions.

Asymptotically motivated inference usually requires that estimates of model parameters be unbiased, which poses a problem for `black box' machine learning predictors that typically trade off bias and variance to produce predictions with low mean squared error. A long literature in semi-parametric inference (\textit{e.g.,} \citet{robins1994estimation}) has worked to remedy these issues, culminating in a large, recent econometrics literature on debiased machine learning (\textit{e.g.,} \citet{chernozhukov2018double, chernozhukov2022automatic}).

There are many parallels between the debiased machine learning literature in econometrics and a literature on 'prediction powered inference' in the deep learning space (\textit{e.g.,} \cite{angelopoulos2023prediction, zrnic2023cross}). Broadly speaking, imputing structured characteristics from unstructured data (the focus of the deep learning literature) and causal inference (the focus of the econometrics literature) are special cases of a more general problem of imputing missing data. In causal inference, potential outcomes are missing, whereas in many deep learning prediction applications, low-dimensional structured characteristics are missing because it is prohibitively costly to extract them from high-dimensional unstructured data manually. The prediction powered inference literature examines how deep learning predictions can be debiased using a high-quality auxiliary sample of ground truth labels for the population of interest. This information is used to measure the bias induced by imputation, which is then corrected, ultimately allowing the researcher to perform valid inference without sacrificing the information available from using a model pre-trained on a larger dataset that makes biased predictions. The deep learning model is treated as a black box. One can show that the prediction powered inference framework is equivalent to debiased machine learning.

\section{Reusability and Reproducibility in Deep Learning} \label{Sec:OpenDL}

Deep learning has been built to a remarkable degree upon open science and open data, although recent years have seen a pronounced shift towards proprietary models and data as the commercial potential of the technology has become increasingly clear. Nevertheless, the amount of open resources is staggering, and deep learning would not have made the strides it has without widespread sharing of models and datasets. Given the centrality of transfer learning and massive-scale pre-training, the field as we know it would not exist without open science. 

The more economics can create an open science culture around big data, whenever data privacy concerns allow, the more we can benefit as a profession from the positive externalities of transfer learning. For example, deep learning researchers are often incentivized to share their code on GitHub as soon as possible, as a way of staking claim to their contribution, or to release a dataset as soon as it is constructed so that a fast-moving literature will use it for longer. Moreover, publication venues in deep learning often require compliance with agreed-upon metadata and ethical frameworks for data and code release \citep{gebru2021datasheets, mlcommons_croissant, mitchell2019model, 10.1145/3287560.3287596}. While I would not advocate that economics wholesale adopt these standards, it is worth considering whether there are standards for model and dataset release that could facilitate reproducibility and reusability of deep neural models in economics.

The largest hub for deep learning models and data is Hugging Face. A wealth of language models and text data can be found there, some examples of which are examined in the demo notebooks linked through EconDL. Hugging Face recently acquired timm, a central repository for vision models, making Hugging Face a one-stop shop for many language and vision tasks. \

\section{Classifiers}
\label{Sec:classification}

Having provided an introduction to deep learning, I now turn to applications. Classification is frequently integral to economic analyses. In the era of big data, a researcher may first need to extract relevant data using classification. For instance, they might start with a massive-scale corpus of news, social media posts, earnings calls, or legislative records and need to extract only coverage about interest rates, immigration, or higher education out of millions or even billions of texts in the full corpus. This much more limited corpus is then used to extract the measure(s) to be used in some downstream causal estimating equation. While this step often receives scant attention, biased classification will result in selection bias into the sample used in the downstream causal estimating equation, potentially significantly biasing the conclusions. Alternatively, a researcher might impute structured data—\textit{e.g.,} geographic locations mentioned in texts, their sentiments or topics, or what type of objects appear in a satellite image—using classification.

This section first introduces classifiers (Section \ref{ssIntroClass}), as well as describing the use of generative AI for classification (Section \ref{ssec:genAI}).
Then, it introduces \textit{sequence classification}, in which a class label is imputed for a sequence of text: \textit{e.g.,} a sentence, paragraph, or document (Section \ref{ssTopicClassification}). It compares the performance of custom-trained classifiers to generative AI on 19 different text classification tasks. 
Classification can also be applied to individual terms in a text (Section \ref{Ssec:NER}).
Finally, a classifier can be used to compare texts to each other (Section \ref{Ssec:cross-encoder}).
I focus on text classification for ease of exposition. Image classification--of a full image or pixels or objects within an image--is analogous, using a CNN or vision transformer rather than a language transformer. It is covered in depth in the EconDL knowledge base post titled `Convolutional Neural Networks.'

\subsection{An Introduction to classifiers} \label{ssIntroClass}

In traditional classification, a neural network predicts a score for each of $N$ classes, and the input is assigned the class with the highest score. For those unfamiliar with classifiers, \citet{3blue1brown} provides an excellent graphical treatment of classification in the context of classifying images of digits.

Recall the analogy that neural networks are like Legos. Central to the power of transformer models is the ability to use the same pre-trained language model as the backbone for a wide variety of classification tasks. This is illustrated in Figure \ref{fig:bert}. 
A transformer language model produces a vector representation for each token (word or sub-word) in its input, as well as the \texttt{<cls>} representation that summarizes the entire input text. 
The text sequence can be classified by adding a classifier head to the \texttt{<cls>} representation (panel a). The classifier is a feedforward neural network that aggregates the nodes in the \texttt{<cls>} vector into a score for each class using learned weights.
As shown in Figure \ref{fig:bert}, panel c, individual tokens can likewise be classified by adding classifier heads to their vector representations. Alternatively, two texts can be jointly embedded and then a classifier can be added to the \texttt{<cls>} representation to classify the relationship between them (panel b).

Training a classifier is one of the most straightforward tasks in deep learning. The open-source package \lt can be used to train text sequence classifiers, with a demo available via EconDL. While the base transformer language model could be frozen when training a classifier, and various layers of the transformer could be used as inputs to the classifier, typically all parameters are allowed to update, with the classifier layer attached to the final layer of the transformer.
Classifiers can be binary (two classes), multiclass (more than two classes, only one of which is positive), or multi-label (multiple classes can be positive). 

Classifier training is a supervised task, and the model must see a sufficient number of examples from each class during training in order to perform well on unlabeled data. When creating labels for classification, the labeled data should be relatively balanced across classes (\textit{e.g.,} positive and negative samples in the case of binary classification).

To train a classifier, we also need an appropriate loss function. The two most common losses for classification are Support Vector Machine (SVM) loss, also referred to as hinge loss, and cross-entropy loss.

Given a sample with true label \( y_i \) and the score vector \( \mathbf{p} \) for the class scores produced by the neural network, the SVM loss is:

\begin{equation}
L_i=\sum_{j \neq y_i} \max (0, p_j-p_{y_i}+1)
\end{equation}

The loss sums over the incorrect classes, imposing a penalty if the score of the correct class is not at least some threshold amount above the score(s) for the incorrect class(es). The threshold can be set to one without loss of generality, as it just scales the learned weights.

Cross-entropy loss measures the dissimilarity between the predicted score distribution and the true distribution. Consider a neural network used for a multi-class classification problem with \( C \) classes. Let \( y \) be the true label of a sample, represented as a one-hot encoded vector. For a sample belonging to class \( i \), \( y_i = 1 \) and \( y_j = 0 \) for \( j \neq i \). Let \( \mathbf{z} \) be the raw numbers (often referred to as logits) produced by the neural network for that sample. A classification layer will produce one score for each class. The predicted score of class \( i \), obtained using the softmax function, is:

\begin{equation}
 p_i = \frac{e^{z_i}}{\sum_{j=1}^{C} e^{z_j}} 
\end{equation}

While the class scores are frequently referred to in the literature as `probabilities', they are not probabilities in a statistical sense. How peaky they are will depend on the regularization of the neural network. 

The cross-entropy loss between the true label and the predicted distribution is:
\begin{equation} \text{CE}(y, p) = - \sum_{i=1}^{C} y_i \log(p_i) \end{equation}

As \( y \) are one hot vectors, this simplifies to: 

\begin{equation}
 \text{CE}(y, p) = - \log(p_{\text{true class}})
\end{equation}
where \( p_{\text{true class}} \) is the predicted probability of the correct class.

With SVM loss, once the score for the correct class surpasses a threshold, it is indifferent to elevating the score of the correct class further. On the other hand, the cross-entropy loss pushes the correct class towards 1. This means that during the early stages of training, accuracy might increase suddenly without a significant change in the loss. In real-world scenarios, both losses typically yield comparable results.

Binary classifiers are evaluated using the F1 score, a metric that combines recall (true positives divided by true positives plus false negatives) and precision (true positives divided by true positives plus false positives).

\begin{equation}
F1 = 2 \times \frac{\text{precision} \times \text{recall}}{\text{precision} + \text{recall}}
\end{equation}

Perfect precision and recall yield an F1 score of 1, whereas the worst score is 0. F1 is a harmonic mean of precision and recall and thus tends to be closer to the smaller of these two metrics. If either the precision or the recall is low, the F1 score will also be low. F1 is preferred over accuracy because if classes are imbalanced, accuracy can be high simply by always predicting the majority class. 

When using a classifier with more than two classes, the F1 score for each class can be calculated individually and then combined in different ways. One approach is Macro F1, where each class’s F1 score is averaged equally. Another method is Weighted F1, where each class’s F1 score is weighted by the number of true instances in that class. A third option is Micro F1, where global true positives, false positives, and false negatives are aggregated across all classes, and F1 is computed as in binary classification. Macro F1 is best suited when equal performance across classes is desired, while weighted or micro F1 are better choices when the dataset is imbalanced and performance on the more frequent classes is most important.

\subsection{Generative AI for classification} \label{ssec:genAI}

Large generative AI models like GPT, Claude, or Llama (commonly referred to in the literature as \textit{foundation models}) use a decoder transformer architecture (Section \ref{arch:transformerllm}) to autoregressively generate text given a prompt. In practice, they might also be connected to an external database (\textit{e.g.,} the internet) via a retrieval-augmented language modeling (RALM) setup (the post on retrieval in EconDL contains more information about RALM). In a fundamental sense, these models are performing classification, autoregressively predicting the most likely next token in a discrete vocabulary at each time step. By default, models like GPT are stochastic; they predict the next token from a distribution of the most probable tokens.\footnote{At the time of writing, setting \texttt{top\_p} to 0 makes GPT deterministic.}

To perform classification tasks using generative AI, the user needs to prompt the model. Prompting a generative language model is, in many ways, less straightforward than tuning a classifier via gradient descent, as the space of discrete prompts is infinite and prompting has generated a large and unwieldy literature.

A few clear insights are though worth emphasizing. Centrally, prompt tuning should be done on a validation set, never on the test set used to evaluate performance. The latter may overfit the prompt to the idiosyncrasies of the test set, making performance on it unrepresentative of performance on the unlabeled data.

A literature on chain-of-thought prompting suggests breaking tasks down into simple steps, making them more digestible \citep{wei2022chain}. This aligns closely with my experience, where simple prompts work much better than lengthier and more detailed ones. If a problem requires lengthy instructions, try to break it down into multiple problems, prompting the model at each step. There is also literature on demonstrating tasks for generative LLMs (\textit{e.g.,} \citet{khattab2022demonstrate}). Whether this is useful will depend on the nature of your task, and I recommend checking whether demonstration helps using a validation set. \citet{liu2023pre} provide a review of prompt engineering. For autoregressive models like GPT, they recommend \textit{prefix} prompts; \textit{e.g.,} ``I love this class. What’s the sentiment of this review?'' This contrasts with cloze prompts: ``I love this class, it is a $[z]$ class.''

This paper examines the performance of GPT-3.5 and GPT-4 on topic classification of historical newspaper articles. Over the past year, I have performed this exercise with older and newer models, as well as GPT-4o and GPT-4 Turbo. GPT-4 and GPT-4o perform similarly, edging out Turbo. GPT-4o mini performs worse. I haven't seen any systematic improvements with new releases. Your mileage may vary, with the new releases that will no doubt be available by the time this article is published.

I also examined two other leading AI models, from Anthropic: Claude Haiku and Claude Opus. These led to significantly worse performance (with F1 scores typically 10-40 points lower than GPT) and are not reported due to space constraints. The drivers of this lower performance are twofold. First, Claude refused to produce an output for texts that it assessed as harmful. A distinguishing feature of Claude is its `Constitutional AI' framework, which sets out certain ethical principles (\textit{e.g.,} harmlessness necessitates that responses should be peaceful, ethical, and avoid content that might be considered offensive in non-Western cultures). Some articles on past conflicts (mostly objectively reporting on events) and a range of other topics (\textit{e.g.,} content about the introduction of contraception written in the 1960s) were considered harmful. Moreover, Claude didn't always generate the desired Yes/No format, making it impossible to extract whether the article was on-topic. Neither of these behaviors arose with GPT. Perhaps this could be fixed with more prompt tuning, or it could change with a new release. 

Regardless, with a classifier it is often fairly straightforward to interpret why it makes particular errors and how to fix them (by adding more training data for the types of instances it is confusing), whereas GenAI at present can feel more like a black box. There is more going on under the hood, with models trained with reinforcement learning to produce responses deemed desirable by the commercial entities training the models. This may or may not pose a problem for a given academic application and underscores the importance of rigorous evaluation with a test set.

The pros of generative AI for classification are: 1) startup costs are low, requiring minimal programming expertise or understanding of what is going on under the hood, and 2) it can be used \textit{zero-shot} (without the user providing training data), whereas tuning a classifier requires training data. We will see that, if anything, custom-trained classifiers tend to have a performance edge on text classification, but straightforward tasks can be performed well with generative AI. This is consistent with the broad consensus that human-level performance can generally be attained on supervised tasks given adequate, high-quality training data.

This brings us to potential disadvantages of generative AI. Using a large model behind an API does not provide the same fine-grained control as training a classifier. While models such as GPT allow users to expose the model to empirical examples, for the experiments in this paper, this did not lead to improvements in performance. It is not fully understood how demonstration through prompting conditions these models, whereas it is clear how providing training examples updates a classifier through gradient descent. Not all models learn equally when exposed to the same amount of training data (as demonstrated empirically in Section \ref{sec:OtherMethods}), and lightweight models like those used with customized classifiers tend to update very efficiently. A custom classifier also has an edge when it comes to interpretability and reproducibility. Results from a commercial API may no longer be reproducible if a model is deprecated, and as discussed above, commercial GenAI models can be more of a black box. These concerns could be mitigated by using an open-source foundation model, such as Meta AI's Llama \citep{touvron2023llama}. The startup costs and hardware requirements, however, negate the ease-of-use advantage. Finally, classifiers using a lightweight backbone such as RoBERTa \citep{liu2019roberta} are very cheap to deploy over a massive number of texts, whereas commercial models at present can be expensive for large-scale problems. This may change if competition increases and research on cheap deployment advances.

To decide whether a classifier or generative AI is most suitable for a task, I would recommend first doing a back-of-the-envelope calculation to ensure that generative AI is within budget. If so, create test and validation sets, tune prompts, and evaluate its performance. If performance is not adequate, a training set to tune a custom classifier will be necessary. If the user knows ex ante that guaranteeing reproducibility is imperative, that there is substantial domain shift from web texts, or that the task requires fine-grained control, they might go straight to training a custom classifier. Data privacy requirements can add an additional layer to consider for those working with confidential data.

\subsection{Sequence classification} \label{ssTopicClassification}

Economists might wish to impute a variety of structured information at the level of a text: \textit{e.g.,} its topic, the type of content it contains, or its sentiment. To illustrate text sequence classification, this section trains 19 different binary topic classifiers, applied to massive-scale databases of historical news \citep{dell2023american, silcock2024newswire}, and compares them to generative AI. It would have been very difficult for annotators to keep 19 different topic definitions in mind to create multi-class labels; hence, binary classification is used. Binary classifiers cannot be combined into a multi-label classifier, as negatives for one topic may be positives for another. Keeping prompts simple for generative AI also suggests binary classes.
 
The annotated data were congruence-labeled by highly skilled annotators, with discrepancies resolved by hand. Congruence labeling ensured high-quality data and facilitated the development of a well-specified definition. For example, in the case of the crime classifier, annotators disagreed on whether articles about Watergate should be considered on-topic, and a zero-shot model showed a massive spike in crime coverage in 1974 due to Watergate. While potentially a reasonable definition, I did not want crime coverage to be skewed by political scandals (which receive massive coverage), and the classifier quickly learned this with a modest number of labels.

A frequent question is how many labels are required. This will vary. Topics that are more diverse or require learning a more complex definition will require more labels. Topics that were seen many times in the pre-training of the language model may require fewer labels. Fortunately, training a classifier is compute-efficient. If, after the first round of training, results are unsatisfactory, it is straightforward to add more labels and retrain. An error analysis may give a sense of what types of texts require more training examples. Since labeling is costly, I recommend starting with fewer labels and adding more if needed.

Table \ref{tab:topic_classification} provides the split statistics for the various topic classification tasks examined in this section.\footnote{The politics classifier is taken from a published paper \citep{dell2023american} with different aims, and hence the split shares and overall number of annotations differ somewhat.} Labeled data are randomly split into training, validation, and test data. Validation data are used to select hyperparameters, select the model checkpoint (when to stop training), and to tune prompts, whereas the test data were used only to compute Table \ref{tab:topic_classification}. The prompts for Table \ref{tab:topic_classification} are listed in the appendix (Table \ref{tab:prompts-1}.

The classifiers were trained with \lt, which supports using any base language model available on Hugging Face. We used DistilRoBERTa (82M parameters) \citep{sanh2019distilroberta} and RoBERTa large (335M parameters). RoBERTa \citep{liu2019roberta} is a widely used, improved version of BERT. Distilled language models are smaller models that are trained to match the performance of a larger model. The distilled version runs faster but often with a performance loss. We used a consistent set of hyperparameters across classification tasks, which appear to work well more generally (a learning rate of $1e-6$ or $1e-5$ and a batch size of eight).

\begin{table*}[t]
\centering
\caption{F1 on test set and summary of train-test split}
\vspace{0.5ex}
\resizebox{\linewidth}{!}{\begin{tabular}{|l|cc>{\centering\arraybackslash}p{0.12\linewidth}>{\centering\arraybackslash}p{0.1\linewidth}>{\centering\arraybackslash}p{0.1\linewidth}|ccc|}
\hline
\rule{0pt}{3ex}
\textbf{Topic} & \multicolumn{5}{c|}{\textbf{F1 on test set}} & \multicolumn{3}{c|}{\textbf{\# of labels}} \\[2ex]
\textbf{} & GPT-3.5 & GPT-4 & GPT-3.5 Trained Model\textsuperscript{\textdagger} & Distil RoBERTa & RoBERTa Large & Train & Eval & Test \\[3ex]
\hline
advice & 0.72 & 0.85 & 0.55 & 0.87 & \textbf{0.97} & 319 & 68 & 68 \\
antitrust & 0.85 & \textbf{0.94} & 0.84 & 0.92 & \textbf{0.94} & 329 & 70 & 70 \\
bible & 0.52 & 0.81 & 0.10 & 0.85 & \textbf{0.87} & 314 & 67 & 67 \\
civil rights & 0.59 & \textbf{0.87} & 0.54 & 0.85 & \textbf{0.87} & 943 & 202 & 202 \\
contraception & 0.75 & 0.91 & 0.72 & 0.88 & \textbf{0.97} & 597 & 127 & 127 \\
crime & 0.85 & 0.80 & 0.85 & 0.85 & \textbf{0.90} & 463 & 98 & 98 \\
horoscope & \textbf{1.00} & \textbf{1.00} & 0.92 & 0.96 & \textbf{1.00} & 288 & 61 & 61 \\
labor movement & 0.77 & 0.89 & 0.79 & \textbf{0.94} & 0.91 & 253 & 54 & 54 \\
obituaries & 0.98 & \textbf{1.00} & \textbf{1.00} & 0.96 & \textbf{1.00} & 272 & 57 & 57 \\
pesticide & 0.58 & 0.91 & 0.71 & 0.89 & \textbf{0.98} & 873 & 187 & 187 \\
polio vaccine & 0.92 & \textbf{0.99} & 0.94 & 0.97 & 0.97 & 350 & 74 & 74 \\
politics & 0.67\textsuperscript{*} & 0.62\textsuperscript{*} & 0.74 & \textbf{0.86} & 0.85 & 2,418 & 498 & 1,473 \\
protests & 0.74 & 0.81 & 0.79 & \textbf{0.91} & 0.90 & 351 & 75 & 75 \\
Red Scare & 0.81 & 0.86 & 0.79 & 0.90 & \textbf{0.91} & 1,852 & 396 & 396 \\
schedules & 0.79 & 0.95 & 0.81 & 0.95 & \textbf{0.96} & 346 & 74 & 74 \\
sports & 0.80 & 0.92 & 0.88 & \textbf{0.94} & \textbf{0.94} & 339 & 72 & 72 \\
Vietnam War & 0.91 & 0.94 & 0.98 & 0.98 & \textbf{0.99} & 738 & 157 & 157 \\
weather & 0.94 & 0.92 & 0.94 & \textbf{0.95} & \textbf{0.95} & 569 & 57 & 57 \\
World War I & 0.72 & 0.74 & 0.51 & 0.89 & \textbf{0.92} & 690 & 164 & 192 \\
\hline
\addlinespace[1ex]
\end{tabular}}
\begin{tablenotes}
\textdagger This column reports the F1 for trained models (based on either DistilRoBERTa or RoBERTa-Large, whichever works better) using labels generated by GPT-3.5. \\
\textsuperscript{*} The results with asterisks were produced on a random sample of 500 out of total 1,473 articles in the test set.
\end{tablenotes}
\label{tab:topic_classification}
\end{table*}

In most cases—across a diversity of topics—the tuned classifier tends to outperform or equal the performance of GPT, though for the more straightforward tasks, GPT's performance can be very good, particularly in the case of GPT-4. The training data used to produce these classifiers are high quality. With lower-quality labels, such as those created with online annotation platforms where quality is notoriously poor, a custom-trained classifier may well be consistently worse than GPT. These comparisons may also change in the future with further innovation in open source and commercial models.

More generally, generative AI performs best on straightforward topics that it was likely extensively exposed to during pre-training. The further the domain shifts from training data—primarily modern web texts—the more performance deteriorates. For horoscopes, obituaries, and articles on the polio vaccine—all extremely straightforward—GPT performs nearly perfectly (as do the classifiers). However, there are also topics for which GPT performs poorly; for example, politics, a topic that is challenging because it is broad and diverse, with content drawn from the late 19th and early 20th centuries and including both local and national politics. World War I has an F1 score from both GPT models in the low 70s, much worse than the Vietnam War, which likely has greater representation in the training corpus. Moreover, language has changed more since World War I, translating into greater domain shift. Yet with minimal labels, the RoBERTa classifiers can adjust to this domain shift.

\begin{figure*}[ht]
    \centering
    \includegraphics[width=\linewidth]{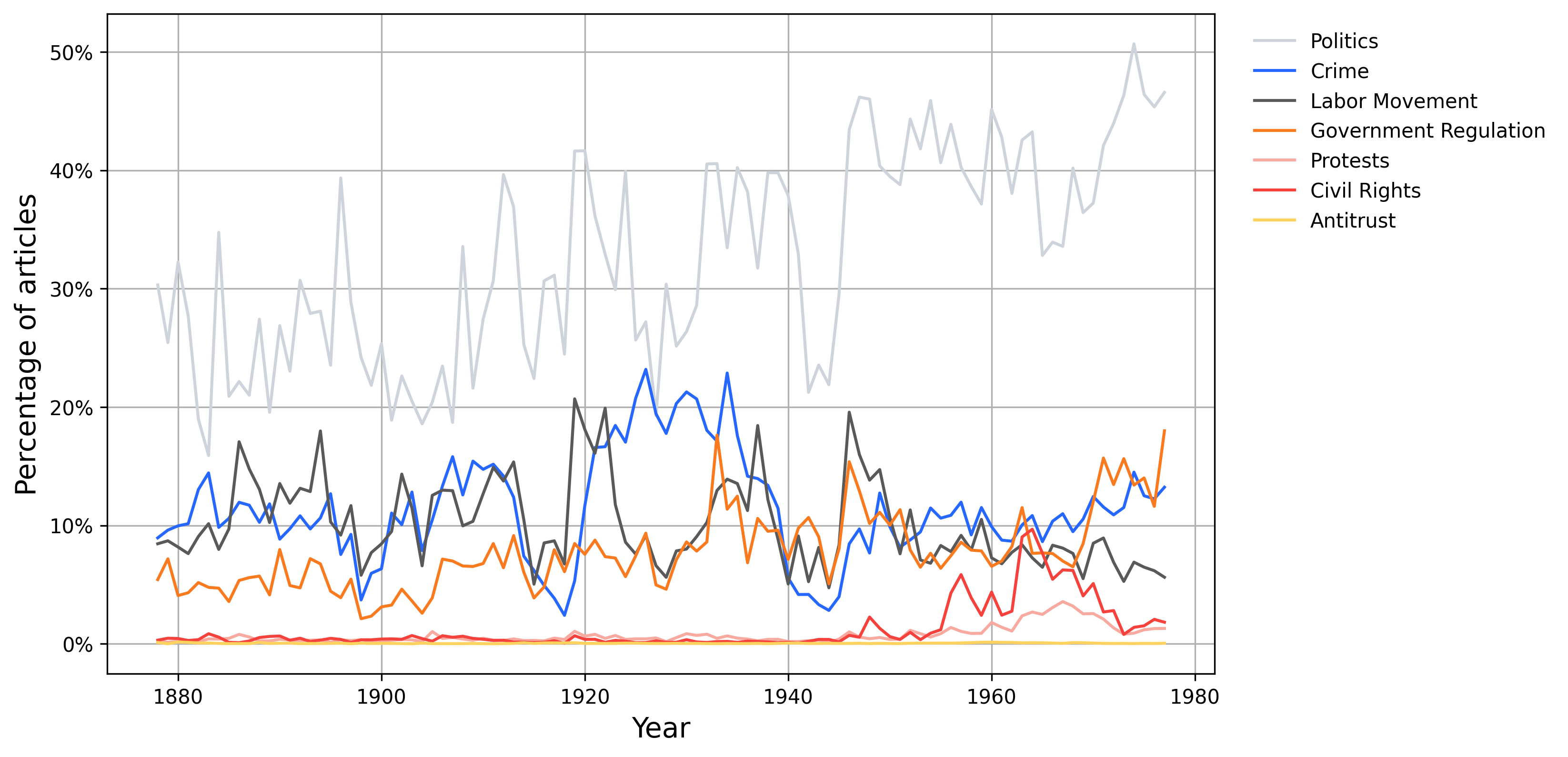}
    \caption{\raggedright Share of newswire articles with a given binary topic tag.}
    \label{fig:topics}
\end{figure*}

At present, GPT is likely to be beyond the budget of most social science researchers for massive corpora, though I do not cite figures here as prices fluctuate and could change considerably depending on competition and technological advancement. Training a RoBERTa classifier on the number of labels shown here is very cheap (at the time of writing, it could be done within minutes on a \$9.99/month Google Colab plan or a mid-range Nvidia GPU card). I have also had students, with patience, train similar models on a laptop, though getting access to a decent GPU through cloud compute or dedicated hardware is preferable. Deploying the classifiers, even across millions of articles, is also cheap, and can be done either with cloud CPUs or on a mid-range GPU card in hours. One can circumvent the expense of generative AI by training a classifier on labels predicted by GPT. Table \ref{tab:topic_classification} shows this can work when GPT produces very high-quality labels. However, where GPT performs less well, training on noisy data can magnify errors.

Figure \ref{fig:topics}, drawn from \citet{silcock2024newswire}, takes binary classifiers that apply across time and deploys them to a dataset of 2.7 million unique newswire articles published between 1878 and 1977. Various trends are evident, such as the Prohibition-related crime coverage spike in the 1920s or the surge of Civil Rights and protests coverage in the 1960s.

It is worth saying a word about how neural methods compare to sparse methods for text classification, which (with varying degrees of sophistication) rely keywords. For example, one common sparse method is TF-IDF: Term Frequency (TF) is the raw count of term \( t \) in document \( d \). Inverse Document Frequency (IDF) measures the importance of a term in the entire corpus. If a word appears in many documents, it's not a good identifier of a given document. The IDF of a term \( t \) for a corpus \( D \) is given by $\log \frac{\text{Total number of docs in } D}{\text{Number of docs with term } t \text{ in } D}$. The TF-IDF score for a term is simply the product of its TF and IDF scores. The higher the TF-IDF score, the more important a term is to a specific document relative to the context of the entire corpus. To rank documents from a corpus based on their similarity to a query using TF-IDF, each document and query is represented as a sparse, high-dimensional vector, with each dimension corresponding to a unique term from the corpus. The weight of each term in the vector is its TF-IDF score. The angle between any two vectors captures the similarity between the texts. One can think of this as akin to a keyword search that downweights terms that appear across the corpus.

We refer to methods like TF-IDF as sparse because each term in the corpus forms a dimension in the vector space. Most terms in the vocabulary will not be present in a single document, leading most entries in the term vector to be zero.

Sparse methods are useful when exact term overlap is highly informative. However, relying on term overlap is often a major shortcoming, as language is complex. There are many ways to say the same thing, and the same term can have different meanings. Moreover, noise (\textit{e.g.,} typos, OCR errors, abbreviations) is ubiquitous. Semantics vary across time and space, as do many omitted variables. This could result in correlation between prediction error and the error term in the causal estimating equation where the keyword predictions are used. Moreover, while terms can be mined, more often they are simply chosen, creating a researcher degrees of freedom problem.

Neural methods address these shortcomings by using a large language model to map texts to a dense vector representation, \textit{e.g,} a 768-dimensional vector composed of non-zero terms. The dimensionality of this vector depends on the base language model. The pre-trained language model is imbued with language understanding, and hence dense methods account for contextual and semantic similarities. This allows them to generalize over synonyms and semantically similar phrases, and to be more robust to other noise. 

\cite{dell2023american} compare the neural classifier for politics, shown above, to mined keywords as well as keywords suggested by ChatGPT. Neural methods lead to significantly more accurate predictions. 

\subsection{Token classification} \label{Ssec:NER}

Researchers may need to extract information about individual terms in a text, rather than the text as a whole. 
This problem is analogous to sequence classification, except that classifier heads are added to the representations for each token in the final layer of the transformer, rather than only to the \texttt{<cls>} representation (Figure \ref{fig:bert}, panel c).

This section develops an example of token classification, named entity recognition (NER), which detects named entities in texts.
These entities can be defined however the researcher desires, as long as a clear, consistent definition and sufficient labels for training exist. 
For example, the researcher may want to identify locations referred to in social media posts. Alternatively, a researcher extracting family relationships from obituary data might wish to tag the relationships of individuals to the deceased (child, parent, sibling, officiant, etc). Or, a researcher wishing to convert biographical texts into a structured dataset might tag birthplace, mother, father, university, spouse, and employer.

\input{figures/NER}

NER is a classic task that has generated a very large literature, and there are many open-source pre-trained models and datasets on Hugging Face. In this literature, entity classes typically include person, location, and organization. CoNLL is a prominent benchmark \citep{sang2003introduction} used to pre-train various models on Hugging Face. 
WNUT is another \citep{nguyen2020wnut}, which focuses on noisy user-generated texts (tweets). A researcher will need to create labels if they depart from the entity types emphasized by the benchmarks. NER typically uses BIO labeling - the first token in an entity is labeled B (for begin), the following tokens are labeled I (for interior), and tokens that are not of interest are labeled O. If the entity types of interest are people (P) and locations (L), tags would be B-P, B-L, I-P, I-L, and O.

Figure \ref{fig:ner} presents results from applying NER to historical newswire articles, plotting the shares of over 27 million entities that fall into four categories: person, location, organization, and miscellaneous.  
The results are sensible, for example showing a spike in location and miscellaneous named entities (\textit{e.g.,} aircraft names) during World War II.

One can also ask generative AI to recognize entities in text and convert the output into a table. 
As with sequence classification, researchers can test whether performance is adequate for their needs by constructing representative validation and test sets. 

\subsection{Relationships between texts} \label{Ssec:cross-encoder}

There are a variety of contexts in which we would like to measure whether two texts are related in some pre-specified way. For example, we could phrase topic classification as a task in which we would like to classify whether one statement entails another: does the article text entail the statement ``this article is about monetary policy''? Alternatively, are two texts noisy duplicates of each other? Do they take the same stance on a political issue? Does one follow the other?

Figure \ref{fig:cross_bi} illustrates two approaches to comparing texts. The top shows a \textit{cross-encoder}: two texts are concatenated, with a \texttt{<sep>} token between them. These texts are jointly passed into the transformer, and a classifier head classifies how they are related. This approach allows for full cross-attention between all tokens in each of the texts. The bottom figure illustrates a \textit{bi-encoder} approach: texts are embedded separately, and then we compute the similarity between them using some distance metric, such as cosine similarity. Bi-encoders are very useful, and we elaborate on them in Section \ref{sec:metric}.

\input{figures/cross_encoder}

A cross-encoder allows for full cross-attention between terms. Because they are jointly embedded, the terms in the texts being compared attend flexibly to each other when creating representations. In contrast, a bi-encoder computes a single representation for each text separately, and then these representations are compared. Hence, cross-encoders tend to have higher accuracy. However, they also have significant drawbacks. Most centrally, if we need to compare $M$ texts to $N$ other texts, this would require embedding $M \times N$ texts. This quadratic cost quickly becomes infeasible, since each text is passed through a neural network with hundreds of millions of parameters.\footnote{Moreover, if we ask a cross-encoder to classify if $a$ is the same as $b$, $b$ is the same as $c$, and $a$ is the same as $c$, intransitivities may result.} In contrast, for the bi-encoder to compare $M$ texts to $N$ other texts, only $M+N$ embeddings are required, making this approach highly scalable. To get the best of both worlds, the literature often uses a bi-encoder to get the $n$ most similar texts to a query text (for some small $n$), and then re-ranks these matches with a cross-encoder. 

\section{Embedding models} \label{sec:metric}

To estimate a classifier, the classes must be specified ex ante, since the number of classes determines the number of parameters in the neural network, and classes must be seen in training. Prompting generative AI for classification tasks also entails specifying what the classes of interest are. However, there are a variety of problems where the classes are not known ex ante or where the researcher would like to add new classes later without having to retrain the model. Moreover, if the number of classes is large, it can become computationally intractable to compute the softmax over all the classes for the loss function.

These common scenarios can be addressed by working directly with the embeddings from the final layer of the transformer or CNN, rather than estimating an additional neural network layer (the classifier) that maps embeddings to class scores. This avoids the pre-specification of classes. Moreover, vector similarity calculations are highly optimized, allowing for problems at the scale of millions or even billions of classes.

This section introduces a variety of applications that can be approached with embedding models. Record linkage (\textit{e.g.,} of individuals, firms, locations, or products across datasets)  is a common task that can be framed as a classification problem with many classes (\textit{e.g.,} each individual, firm, etc.). It is particularly amenable to embedding methods (Section \ref{Ssec:LinkLLM}).
These methods can handle settings---like multilingual linkage or linkage with multiple, noisy text descriptions---that are very difficult to tackle with traditional string matching methods.
Similar methods can be used to link mentions of individuals, firms, etc. in unstructured texts (\textit{e.g.,} social or print media, firm filings, biographies, government documents) to an external knowledge base such as Wikipedia (Section \ref{Ssec:entity_disambig}).
Tracking the spread of text or images through media is another application of potential  interest, where the classes are often unknown ex ante (Section \ref{Ssec:wireclustering}). 
In some cases, the aim may be exploratory, to uncover the stylized facts in a massive novel text or image dataset, and embedding methods are well-suited to such descriptive analyses. 
Finally, optical character recognition (OCR) can be framed as an image classification task, where the researcher might wish to add characters or words subsequently without retraining the model, suggesting embedding methods (Section \ref{Ssec:OCR}). 

Working directly with embeddings requires distances between vector representations to be meaningful. The geometric properties of pre-trained transformer language models are not well-suited to this task. For example, representations of low-frequency words are pushed outwards on the hypersphere. The sparsity of low-frequency words violates convexity, and the distance between embeddings is correlated with lexical similarity. This leads to poor alignment between the embeddings of semantically similar texts and poor performance when individual term representations are pooled to create an average representation for text sequences \citep{ethayarajh2019contextual, reimers2019sentence}.

Mathematically, the problem is that the embedding space created by a pre-trained transformer model is not \textit{isotropic}, meaning that the representations are not evenly distributed. When embeddings are isotropic, no particular direction is favored. This uniform distribution ensures that the distances between vectors accurately reflect their relationships, making the space more effective for tasks that depend on these distances. Contrastive learning is a widely used method that improves isotropy and is discussed before turning to embedding model applications.

\subsection{Contrastive learning}

Contrastive learning aims to learn similar representations for semantically similar inputs and dissimilar representations for semantically different inputs, where the definition of similarity is given by empirical training examples. The contrastive loss function encourages the model to reduce the distance in embedding space between positive examples (\textit{e.g.,} similar texts or images) and increase the distance between negative examples (\textit{e.g.,} dissimilar texts or images). Contrastive training reduces anisotropy \citep{wang2021understanding}, significantly improves pooled representations of text sequences \citep{reimers2019sentence}, and improves alignment between representations of semantically similar texts.

Contrastive learning follows the bi-encoder setup shown in Figure \ref{fig:cross_bi}. Bi-encoders form a single representation for each instance, by passing it through a transformer and pooling (averaging) term level embeddings.\footnote{Representations are pooled rather than using the \texttt{<cls>} token because this has been shown to have better performance \citep{reimers2019sentence}.} Representations can be compared by computing their vector similarity. In practice cosine similarity is frequently used since representations are on a unit hypersphere. 

There are different options for the loss function depending on the training data. Contrastive loss \citep{chopra2005learning} uses positive and negative pairs, incentivizing positives to have the same representation and negatives to be above a threshold distance apart. A cosine loss uses a continuous measure of the difference between instances. With triplet loss \citep{hermans2017defense}, training data consist of triplets: an anchor, one positive example for the anchor, and one negative example. Embeddings of positives are incentivized to be more similar to the anchor than negatives. With the InfoNCE loss \citep{oord2018representation}, multiple negative examples are compared to a single positive example. Supervised contrastive loss \citep{khosla2020supervised} generalizes InfoNCE to allow for multiple positive and negative examples. More details, including mathematical formulations, are provided in the `Contrastive Learning' post in the EconDL knowledge base.

The metric space created by a neural network should be interpreted based on the contrastive loss function used to train the network. For instance, with contrastive loss, instances in the same class are incentivized to have similar representations, whereas different instances are incentivized to be above a threshold distance apart. Hence, local distances are meaningful, but global distances are not, since being more dissimilar beyond the threshold does not affect the loss.

When contrastively training on paired data, the weights of the two encoders used to embed each of the instances can be the same (a symmetric encoder) or different (an asymmetric encoder, as in \citet{karpukhin2020dense}). Symmetric encoders provide a more parsimonious model, and hence require less data and compute to train. In practice, they can perform well even when two distinct types of instances are being encoded (\textit{e.g.,} search queries and documents that contain their answers).

Selecting informative negative examples is important for contrastive learning. If the negatives are too `easy'—\textit{e.g.,} they are dissimilar in the embedding space of the pre-trained model used to initialize training—little is learned. In some contexts, the researcher can use prior knowledge to select `hard' negatives for training. In other contexts, they can be mined by using a pre-trained model to choose negative examples with similar embeddings. This can work well if the researcher knows ex ante which instances are negative, as would be the case when training on synthetically generated data. When negatives are drawn from unlabeled data, however, mining hard negatives without a human in the loop risks inadvertently selecting positives. Sometimes training on random negatives may be sufficient, although this often requires large batch sizes (\textit{e.g.,} GPU cards with a lot of memory), which is not amenable to academic compute budgets. This section's applications consider how negatives are selected. 

\input{figures/ComparativeAgendas}

Embeddings are also available off-the-shelf. Sentence-BERT \citep{reimers2019sentence} is a prominent and well-supported open-source model. (In this literature, the term `sentence' is used to refer to any text sequence, which could be a phrase, sentence, or entire document.) Moreover, OpenAI sells quite affordable sentence embeddings. 

While there has been considerable interest in developing all-purpose embedding models that can excel zero-shot on any task \citep{cao2023recent}---and a larger model will on average outperform a smaller model zero-shot---fine-tuned lightweight embedding models have important advantages over zero-shot embeddings. Intuitively, embeddings provide a single representation for each text (or image). An off-the-shelf representation will capture lots of different information about the text/image. However, the researcher is typically interested in some narrowly defined aspects of it. Fine-tuning will accentuate the relevant dimensions, creating better separation between classes in embedding space.

Consider the following empirical example. I take the comparative agendas dataset on U.S. legislation \citep{wilkerson2023policy}, which assigns topic tags to congressional bills, and calculate pairwise similarities between the embeddings of the legislative descriptions using three different models: off-the-shelf lightweight S-BERT embeddings (Figure \ref{fig:comparative_agendas}, panel a), OpenAI large embeddings (panel b), and embeddings produced by tuning S-BERT on paired positives and (random) negative bills from a training split of these data (panel c). The blue distribution plots cosine similarities within topics, and the red line plots cosine similarities between topics. Identical representations have a similarity of 1.

With off-the-shelf models, embeddings are indeed more similar within than across topics (\textit{e.g.,} the blue distribution is shifted to the right of the red distribution), but the differences are not stark. SBERT and OpenAI perform similarly. In contrast, once the model has been tuned on target data, embeddings are much more similar within topic than across topic, as the contrastive training accentuates the importance of topic in determining how the language  model maps texts to embedding space. Much of the overlap in distributions comes from edge cases where articles fall between topics or cover multiple themes. Panel (d) uses the model tuned on U.S. bills to compare embeddings of UK Acts of Parliament within and across topics. While there is some domain shift, there continues to be marked separation, showing the fine-tuned model's ability to generalize to similar problems. 

\subsection{Record linkage with structured data} \label{Ssec:LinkLLM}

\begin{figure*}[htb]
    \centering
    \includegraphics[width=\linewidth]{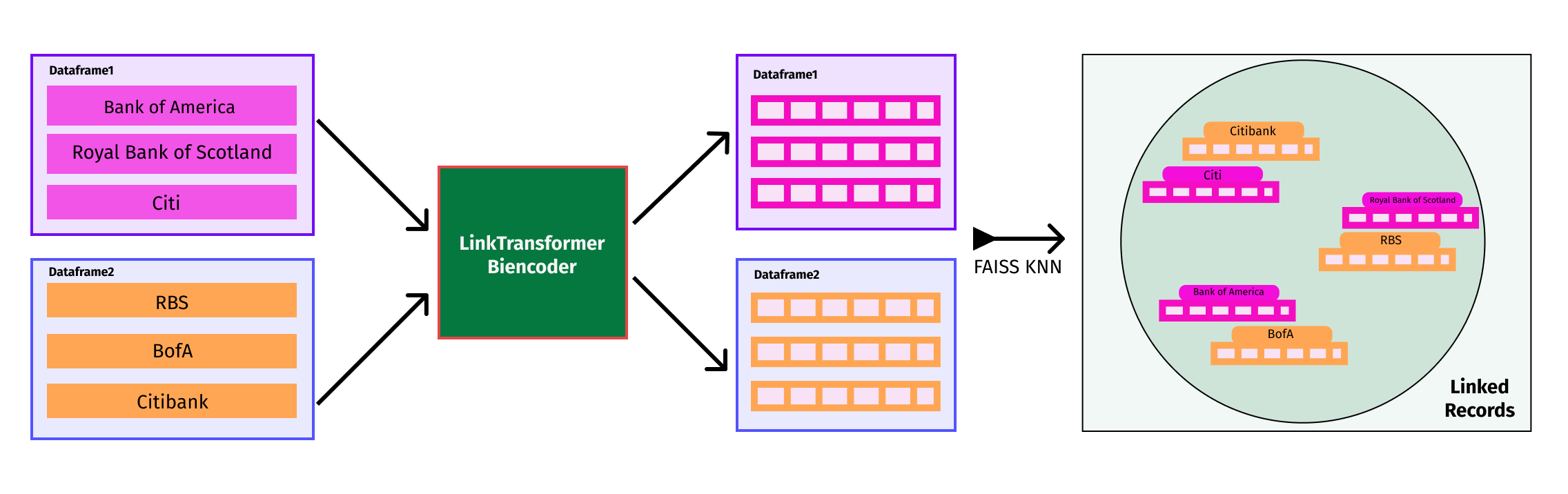}
    \caption{\lt architecture.} 
    \label{fig:LTarch}
    \vspace{-4mm}
  \end{figure*}
  
Record linkage is central to many economic analyses. A researcher might need to link individuals, locations, firms, organizations, product descriptions, or academic papers across datasets. Traditionally, records have been linked using measures like Levenshtein edit distance---which counts the number of character insertions, deletions, and substitutions to convert one string into another---or Jaccard similarity---which computes the similarity between substring n-gram representations of strings. A recent machine learning literature focused on matching across e-commerce datasets shows the promise of transformer LLMs for improving record linkage. Yet at the time of writing, these methods have not yet made widespread inroads in social science, with rule-based methods continuing to overwhelmingly predominate (\textit{e.g.,} see reviews by \citet{binette2022almost, abramitzky2021automated, bailey2020well}). 

To make these methods more accessible, \citet{arora2023linktransformer} designed \lt, a package for using transformer models for record linkage that is geared towards social scientists. The study documents that transformers outperform traditional string matching methods across a variety of tasks and languages, often by a wide margin. Applications include linking 1940 Mexican tariff schedules and linking 1950 Japanese firm-level records using multiple noisy fields, as well as linking modern firms and products across six languages. A multilingual model can link products across languages without the need for translation.

This work was motivated by a variety of projects that I had to abandon in the pre-deep learning era because sparse methods performed poorly and hand-linking was infeasible. As with any predictive task, it is incumbent upon the researcher to evaluate whether performance is acceptable using a test set.

The \texttt{LinkTransformer} model architecture is shown in Figure \ref{fig:LTarch}. The texts that need to be matched are encoded using a transformer language model. For each query, \lt finds the nearest neighbors in the corpus, as measured by the cosine similarity between the embeddings. This is extremely fast, as it uses the highly optimized FAISS (Facebook Artificial Intelligence Similarity Search) backend \citep{johnson2019billion}. \lt returns a ranking as well as the cosine similarity scores, which can be used for 1-1, 1-many, or many-many merges, including no matches (captured when the similarity to the nearest record is below some threshold). 

Just as traditional sparse methods like edit distance return distances between records, \lt computes a distance metric powered by all the semantic knowledge embodied in the pre-trained language model, as well as any additional knowledge gained through contrastive training. For example, ABC Corporation, ABC Co., and ABCC are very similar semantically---and hence nearby in embedding space---since `Co.' and `C' represent `Corporation,' but these strings have a high Levenshtein edit distance. Examples like this are common in record linkage tasks, given the prevalence of abbreviations, different ways to describe the same product or firm name, OCR errors, and typos. Embedding similarity can be used analogously to how researchers use string distance metrics.

\lt seamlessly supports linking with multiple fields, serializing fields by concatenating them with a \texttt{<sep>} token that the package automatically selects to be compatible with the base language model tokenizer. The study provides an example of linking 1950s Japanese firms across different large-scale, noisy databases using the firm name, location, products, shareholders, and banks. This type of linkage problem would be highly convoluted with string matching methods, as fields are noisy (\textit{e.g.,} products are described in different ways across datasets, different subsets of managers and shareholders are listed, etc.). A large language model can handle these challenges with ease because it captures semantic similarity.
 
\lt allows users to employ Sentence Transformer models, OpenAI embeddings, models tuned on the target task, or any transformer language model available on Hugging Face. The general picture that emerges, across 20 diverse linking tasks, is that models custom trained on a modest number of labels tend to perform best, followed by off-the-shelf embeddings from OpenAI, and then off-the-shelf Sentence Transformer models (though there is some variation from task to task). The results are consistent with the discussion of off-the-shelf versus customized embedding models above.

\lt also provides APIs to use transformer large language models for other data processing tasks, \textit{e.g.}, classification, aggregation, and de-duplication, as outlined in tutorial notebooks linked on EconDL. Users can also find a `Train Your Own \lt Model' tutorial, for when customization is necessary. Contrastive training requires both positive and negative pairs (in this case, linked records and distinct records). The user can provide only positives---in which case \lt chooses negatives at random---or can provide both positives and negatives, if hard negatives are available. To promote reusability, reproducibility, and extensibility, models can be shared to the Hugging Face hub with a single line of code. 

When the task is to link scanned documents (as in much of economic history), computer vision may also be useful.
In the vision only record linkage models developed in \citet{arora2023linking}, record linkage is OCR-free, using only the image crops of the firm names to be linked. This is explored in a challenging setting, linking firms between historical Japanese publications where one publication is written horizontally and the other vertically. In general, vision-only linkage leads to reasonably high accuracy. 
However, there are some matches that a vision model cannot resolve because a firm can write its name in different ways. 

Embedding models can combine vision and language transformers, leveraging an understanding of both semantic and visual similarity. \citet{arora2023linking} show that a multimodal model leads to extremely accurate linkage of OCR'ed Japanese firm-level records on customers and suppliers, whereas less accurate string distance metrics produce a different supply chain network that would likely lead to biased downstream economic analyses. 

Combining text and image embeddings is most straightforward when the embedding spaces are aligned. In other words, image and text representations of the same thing need to have similar embeddings; \textit{e.g.,} a picture of an avocado and the text 'avocado' have similar embeddings when an image and text encoder respectively map them to vector space. \citet{arora2023linking} start with a Japanese version of CLIP, which stands for Contrastive Language-Image Pre-training. CLIP is an OpenAI model contrastively trained to align text and image encoders using 400 million image-caption pairs scraped from the web \citep{radford2021learning}. \citet{arora2023linking} do further pre-training on synthetically noised pairs of document crops and their corresponding OCR'ed texts. Pooled text-image representations are then used to link firms.

There are other ways to incorporate visual similarity into linkage that may be helpful when OCR errors are present. Approximate string matching methods count the number of edits (insertions, deletions, and substitutions) needed to transform one string into another \citep{levenshtein1966binary}. In practice, not all string substitutions are equally probable, and efforts to construct lists that vary their costs date back at least to 1918, when Russell and Odell patented Soundex \citep{russell-1918,nara-soundex}, a sound standardization toolkit that accounts for the fact that census enumerators often misspelled names according to their sound. Such methods can significantly improve the accuracy of edit distance linking in the contexts for which they are tailored but are labor-intensive to extend to new settings due to the use of hand-crafted features. This skews research with linked data---necessary to examine many economic questions---towards higher resource settings that are not representative of the diversity of human societies.

\citet{yang2023homoglyphs} develop an extensible, self-supervised method for determining the relative costs of character substitutions in databases created with OCR. OCR often confuses characters with their homoglyphs, which have a similar visual appearance (\textit{e.g.,} '0' and 'O'). \citet{yang2023homoglyphs} augment digital fonts to contrastively learn a metric space where different augmentations of a character (\textit{e.g.,} the same character rendered with different fonts) have similar vector representations. The resulting space can be used, with a reference font, to measure the visual similarity across different characters. Examples of characters and their nearest neighbors in homoglyph space are shown in Figure \ref{fig:char_near}.

\input{figures/char_near}

Using the cosine distance between characters in the homoglyph space as the substitution cost within a Levenshtein edit distance framework \citep{levenshtein1966binary} significantly improves linkage of firms and placenames. The study focuses on CJK, as the extremely large number of characters in this script make it completely infeasible to compute homoglyphs by hand, but shows that the  method is extensible by computing homoglyphs for ancient Chinese characters and for all of Unicode. The broader takeaway is even when traditional methods---like string distance---are preferred, deep learning may provide a way to cheaply extend the methods to novel settings.

\subsection{Linking unstructured data} \label{Ssec:entity_disambig}

There is also a large NLP literature on linking entities mentioned in unstructured texts (\textit{e.g.,} news, social media, etc.), a task referred to as \textit{entity disambiguation}. Linking entity mentions in raw texts (tagged through NER - Section \ref{Ssec:NER}) to Wikipedia or other knowledge bases is useful, because these contain information such as structured biographical data. Whether individuals are in an external knowledge base may itself also be of interest.

Researchers might also wish to coreference entity mentions across documents in a corpus (\textit{e.g.,} find every reference to President John F. Kennedy in historical news). This is referred to as \textit{coreference resolution.}

Despite increasing digitization, historical documents typically lack cross-document identifiers for individuals mentioned in the texts, as well as identifiers from external knowledge bases like Wikipedia, both of which would make it much easier for economists to extract structured data from these sources. 

\cite{arora2024contrastive} develop a bi-encoder embedding model for coreferencing entities within texts and disambiguating them to Wikipedia. The model is contrastively trained on over 190 million entity pairs from Wikipedia. 
Positive pairs come from contexts (paragraphs) in Wikipedia that contain hyperlinks to the same page (for coreference), or from a context and the first paragraph of the relevant entity that it links to (for disambiguation). 
Hard negatives are mined at scale from Wikipedia disambiguation pages, which list entities that have confusable names or aliases. For example, the disambiguation page "John Kennedy" includes John F. Kennedy the president, John Kennedy (Louisiana politician), John F. Kennedy Jr, and a variety of other John Kennedys. Hard negatives sample contexts mentioning John F. Kennedy (\textit{e.g.,} with hyperlinks to John F. Kennedy's page) and pair them with contexts mentioning other entities from the John Kennedy disambiguation page.
Hard negatives from families (\textit{e.g.}, Henry Ford Jr. and Sr.) are over-represented by mining family members from Wikidata. 
It was also necessary to include random negatives, as otherwise the model lost its initial ability to distinguish easy cases, a phenomenon known in the deep learning literature as \textit{catastrophic forgetting}.  

This illustrates how existing knowledge can be mined to create informative negatives for contrastive training.  
More generally, Wikipedia is a useful source of training data (\textit{e.g.,} firm aliases to train \lt models were also taken from Wikidata). 

Entity mentions are disambiguated by embedding their contexts with the disambiguation model and retrieving their nearest Wikipedia neighbor in embedding space. If they are below a threshold cosine similarity to the nearest Wikipedia embedding, they are marked as not in the knowledge base. 
\cite{silcock2024newswire} find that the most mentioned entity in these 100 years is Dwight Eisenhower---edging out Adolf Hitler, Richard Nixon, and Harry Truman---and only 4.7\% of disambiguated entities in newswire articles are women.

\subsection{Classification when categories are unknown} \label{Ssec:wireclustering}

The classes that a researcher would like to impute from unstructured data may be unknown ex ante. This is particularly likely when the aim is to describe the stylized facts in a novel unstructured corpus, but can arise more generally. This section provides several examples drawn from media economics: detecting reproduced article texts and images, classifying the biggest news stories historically, and retrieving historical news stories that are semantically similar to modern ones.

Reproduced content is a fundamental feature of media—both traditional media and in the age of sharing via social media. Media historian Julia \citet{guarneri2017newsprint} writes: ``by the 1910s and 1920s, most of the articles that Americans read in their local papers had either been bought or sold on the national news market... This constructed a broadly understood American `way of life' that would become a touchstone of U.S. domestic politics and international relations throughout the twentieth century.'' Suppose we would like to be able to identify each unique article and image that was sent out over the newswire, measure how widely reproduced it was, and observe which papers reproduced it. This problem is more challenging than it seems. Texts are often heavily abridged and can contain significant OCR errors. Images are often cropped, and the quality can be extremely low.

\citet{silcock332noise} show that deep neural methods can significantly outperform non-neural methods for detecting noisily duplicated texts, with applications to historical newswires, modern news, and patent databases. Training and evaluation data are hand-curated by grouping articles across thousands of digitized local newspapers into groups of articles from the same newswire source. Articles from the same wire source are positives. Articles with high $n$-gram overlap but from different sources—often similar articles from different newswire services or article updates—form hard negatives. Random negatives are also used in training.

A bi-encoder embedding model is contrastively trained such that articles from the same wire article source (regardless of noise and abridgement) have similar vector representations, while articles from different sources (even if about the same underlying story) have different representations. These representations can then be clustered with highly efficient single-linkage clustering to quantify which articles are from the same underlying news wire or syndicate article source and which are from different ones. Community detection is used to break the spurious links that are a potential drawback of single-linkage clustering.\footnote{Other common clustering methods, like hierarchical agglomerative clustering, do not scale well.} As with record linkage, the Sentence-BERT library \citep{reimers2019sentence} was an important resource. The model is initialized with the S-BERT MPNet bi-encoder, a lightweight, high-performing semantic similarity model.

The neural approach outperforms traditional $N$-gram and hashing methods - sparse methods that rely on term overlap to detect noisy duplicates - by a wide margin. As suggested in Section \ref{Ssec:cross-encoder}, modest gains result from adding a re-ranking step that applies a cross-encoder to articles within a threshold bi-encoder distance. 

A strength of embedding methods is their scalability. Clustering dense vector representations requires highly optimized similarity search, as traditional clustering libraries don't scale well. Facebook AI Similarity Search (FAISS) \citep{johnson2019billion}, an open-source library for computing vector similarity, made $10^{14}$ exact similarity comparisons---required to cluster 10 million article representations---on a single GPU card in around 3 hours. This could have been sped up significantly, with only a modest hit to accuracy, by using approximate vector search (with appropriately tuned hyperparameters).

\cite{silcock2024newswire} release 2.7 million unique newswire articles spanning 1878-1977 (the end date is due to copyright law changes). It includes topic tags, named entity tags, disambiguation of individuals to Wikipedia, and the counties where articles ran.

Detecting the noisy reproduction of images is analogous to detecting reproduced texts. Rather than training a language model, a vision model can be contrastively trained to map reproduced versions of the same image to similar vector representations and different images to dissimilar representations. A lightweight CNN works well in practice \citep{howard2019searching}, with little gain from using a much larger ViT. Training data consist mostly of synthetically augmented images—which simulate the noise present in the actual images. When it is possible to simulate realistic synthetic data, this can save considerable annotation expense, though adding a modest number of labeled examples from target data may still offer a performance boost.

\begin{table}[!htb]
    \centering
    \resizebox{\linewidth}{!}{
    \begin{tabular}{ll}
    \textbf{Year} & \textbf{Biggest story} \\
    1885 & Death of General Grant \\
    1886 & Southwest Railroad Strike \\
    1887 & Vatican supports Knights of Labor \\
    1888 & Rail strikes \\
    1889 & Samoan Crisis \\
    1890 & 1893 World's Fair planning \\
    1891 & New Orleans Lynchings \\
    1892 & Homestead Steel Strike \\
    1893 & World's Fair, Chicago \\
    1894 & Wilson–Gorman Tariff Act \\
    1895 & British occupation of Nicaragua \\
    1896 & Bimetallism Movement \\
    1897 & Coal Miners' Strike \\
    1898 & Cuban War of Independence \\
    1899 & Philippine-American War \\
    1900 & Anglo-Boer War \\
    1901 & U.S. Steel Recognition Strike \\
    1902 & Anthracite Coal Strike \\
    1903 & Panama Canal Treaty \\
    1904 & Russo-Japanese War \\
    1905 & Russo-Japanese Peace Process \\
    1906 & Hepburn Railroad Rate Bill \\
    1907 & Mining accidents \\
    1908 & Taft presidential victory \\
    1909 & Race to the North Pole \\
    1910 & Rail strikes \\
    1911 & Canadian Reciprocity Bill \\
    1912 & Republican National Convention \\
    1913 & Underwood-Simmons Tariff Act \\
    1914 & World War I \\
    1915 & World War I \\
    1916 & Pancho Villa Expedition \\
    1917 & World War I \\
    1918 & World War I \\
    1919 & Treaty of Versailles \\
    1920 & Rail strikes \\
    \end{tabular}
    }
    \caption{Biggest news stories.}
    \label{same_story}
\end{table}

Embedding models are well-suited to assessing stylized facts in unstructured data at scale. Deep learning makes it possible to use a variety of novel unstructured datasets in economic research. While our focus is often on using causal estimation to test precisely defined hypotheses, understanding stylized facts is an important first step to formulating these hypotheses.

Consider an application from \citet{dell2023american}, which constructs a historical newspaper dataset consisting of over 430 million historical U.S. newspaper articles. The aim of the exercise is to determine the biggest news stories of each year without knowing what these stories are ex ante. The study contrastively trains a model on data from AllSides, a modern news website that groups news articles from different sources into stories (often with different perspectives on the same event).\footnote{\url{https://www.allsides.com/unbiased-balanced-news}} Grouped stories form positive pairs for training, with the model learning what constitutes the ``same story'' via these empirical examples. 
The trained model is used to embed articles, and stories are formed by clustering. Table \ref{same_story} reports the largest cluster for each year \citep{dell2023american}. Some interesting stylized facts emerge, particularly the extensive coverage of labor movements. If a researcher wanted to create a measure of labor movements to use in a causal estimating equation, they would likely train a classifier with carefully crafted labels to predict which articles are about labor movements. In contrast, this exercise motivates why labor movements are important to study in the first place. 

\citet{franklin2024news} use this model for additional data exploration, first masking out all named entities (people, organizations, locations, and other miscellaneous proper nouns) and then querying the most similar historical news articles to a modern news article query in embedding space. The resulting \ndjv open-source package and website provide a novel tool for exploring parallels in how people have perceived past and present.

\subsection{Optical character recognition} \label{Ssec:OCR} 

Optical character recognition (OCR) is an important task for economists, particularly for economic historians. Documents are extremely diverse in terms of character sets, languages, fonts or handwriting, printing technologies, and artifacts from scanning and aging. Off-the-shelf OCR technology is developed largely for small-scale commercial applications in high-resource languages like English, and the architecture it uses is not well-suited to extending OCR to lower-resource languages and settings, as elaborated in Section \ref{sec:OtherMethods}. OCR quality can deteriorate rapidly when moving away from English and a few other high-resource languages \citep{carlson2023, hegghammer2021ocr}. For example, on printed Japanese documents from the 1950s, the best-performing existing OCR mis-predicts over half of the characters. Poor performance is widespread, spurring a large post-OCR error-correction literature \citep{10.1162/tacl_a_00379, 10.1145/3453476, artidigh20}.

Even in the highest resource settings, off-the-shelf solutions can still fail, especially when accuracy is paramount. This is particularly true when transcribing quantitative data. An OCR error in prose is often straightforward to correct in post-processing or inconsequential. However, with numbers, a similar error (\textit{e.g.,} hallucinating a “1” at the beginning of a number) may severely bias downstream statistical analyses.

Moreover, the scale of document collections to be digitized can be vast. For example, the U.S. National Archives holds approximately 13.28 billion pages of textual records. Bringing big data to economic history requires an OCR technology that is both accurate and cheap to deploy.

If economists, historians, and others rely solely on off-the-shelf commercial technologies, we will end up focusing on economic applications that look a lot like high resource commercial applications (\textit{e.g.,} receipts in English). This is indeed what I've seen over many years of working with students: they are far more likely to abandon projects in lower resource languages because the OCR quality of any existing off-the-shelf solution is poor. This skews economic knowledge towards settings that look more like high-resource commercial applications, which are not representative of the diversity of human societies.

To address these challenges, \citet{carlson2023} develop a novel, open-source OCR architecture, \effocr (\textbf{Eff}icient\textbf{OCR}). \effocr is designed for researchers and archives seeking a sample-efficient, customizable, scalable OCR solution for diverse documents. Deep learning-based object detection methods (Section \ref{Sec:regression}) are used to localize individual characters or words in a document image. Recognition models for characters or words are contrastively trained—largely on augmented digital fonts—to map image crops of the same character or word to similar vector representations, regardless of font and other variations. Different characters or words, even if they have a very similar visual appearance, are mapped further apart. 

A document is transcribed by embedding word or character crops and retrieving their nearest neighbors in an index that embeds crops rendered with a digital font. New characters or words can be added to the index after training (unlike with a classifier), a useful feature for economic historians since idiosyncratic symbols frequently appear in historical document collections.

\input{figures/ca_layouts}

\effocr performs very accurately, even when using lightweight models designed for mobile phones that are cheap to train and deploy. For example, it can provide a sample efficient, highly accurate OCR architecture for historical Japanese documents where all current solutions fail. Its blend of accuracy and efficient runtime also makes it attractive for digitizing massive-scale collections in high-resource languages. \cite{dell2023american} cheaply digitize over 430 million historical newspaper articles from the Library of Congress's Chronicling America collection with \effocr. TrOCR, an open-source solution with similar accuracy, would have cost nearly 50 times more to deploy, and commercial solutions were even more cost prohibitive.  

EconDL links to a demo notebook for training a custom OCR for polytonic (ancient) Greek, using minimal cloud compute. \citet{carlson2023} show that this model outperforms Google Cloud Vision on the target data. The notebook uses the \effocr package \citep{bryan2023}, which allows users to tune their own OCR models and run existing models off-the-shelf.
\effocr does not focus on handwriting; however, the approach would be analogous. Synthetic handwriting generators, \textit{e.g.,} \citet{bhunia2021handwriting}, could provide extensive data for pre-training, analogous to the use of digital fonts.

\section{Regression} \label{Sec:regression}

In machine learning, the term ``regression'' refers to the prediction of continuous outcomes.
Regression using deep neural networks is analogous to classification, except that a regression layer added to a neural network predicts a continuous number(s), rather than a set of class scores. For this reason, our treatment here is brief, focusing on a single application: object detection. 

Object detection problems, as the name suggests, locate objects in an image \citep{ren2017faster, he2017mask, kirillov2019panoptic, cai2019cascade, redmon2016you, ultralytics_2020, carion2020end, liu2021swin}. For example, an economist digitizing firm financial records would need to detect the coordinates of different document objects: \textit{e.g.,} table headers, column and row headers, table cells, footnotes, etc. Alternatively, an economist wishing to measure informality from street view data would need to localize street vendors in an image. For each object, the neural network outputs four continuous numbers (top-x, top-y, height, and width of the box containing the object)---a regression problem---as well as the class of that object (\textit{e.g.,} table header, column header, etc.)---a classification problem.

\input{figures/gcv}

Figure \ref{fig:ca_layouts} shows how object detection methods can be used to localize and classify document layout objects (\textit{e.g.}, articles, headlines, etc.) in scans of historical newspapers, facilitating the creation of structured digital texts that can be analyzed with modern NLP methods. 
In contrast, Figure \ref{gcv} provides an example of how commercial OCR (Google Cloud Vision) reads a newspaper scan like a single column book, failing to detect individual articles, headlines, etc. 
All one can do with these scrambled texts is search for keywords, as is typical in the economic literature using historical newspapers (see \citet{hanlon2022historical} for a review).
Layout detection is also needed to extract structure when digitizing tabular data, as Figure \ref{fig:tk} shows for historical Japanese firm records. 

\input{figures/tk_layouts}

At present, document layout detection typically requires customization. 
An off-the-shelf model may work well if the target task is quite close to what it was fine-tuned for. 
However, in computer vision, the main pre-training dataset is ImageNet, which consists of natural images, such as different breeds of dogs. Models have not been exposed to massive-scale pre-training on documents, and hence tend to suffer from substantial domain shift when applied to different types of documents. This is particularly true with historical documents, which are heterogeneous. 

While there is a foundation model for object localization, Segment Anything by Meta AI \citep{kirillov2023segment}, at the time of writing I have not found it to be very useful for document tasks. It can localize objects in an image, but does not classify them into categories. Moreover, localizations on document images are not particularly accurate at present. 

The open-source package \lp \citep{shen2021layoutparser} lowers the barriers to detecting document layouts with deep learning. The library---which contains a model zoo for off-the-shelf usage and facilitates tuning customized models---is implemented with simple Python APIs. More information can be found on the EconDL resource page. 
The EconDL knowledge base introduces active learning for object detection \citep{shen2022olala}, which chooses instances to label that the model is most uncertain about to economize on labeling costs.

Space constraints preclude delving into the architectures of object detection models, but a detailed treatment is provided in the object detection post in the EconDL knowledge base. Resources for satellite images include \textit{e.g.,} \citet{aleissaee2022transformers, wang2022empirical, bandara2022transformer, fuller2022transfer} and are linked from the page on processing satellite imagery in EconDL.

\input{figures/EffOCR_arch}

\section{Alternative methods} \label{sec:OtherMethods}

This review has focused on a set of methods that---while a mainstay in deep learning---are far from comprehensive. If the reader delves into the deep learning literature, they will find other approaches to the above problems and may wonder why those were not covered. I focus on classifiers and embedding models because they are often sample and computationally efficient, meaning that they learn well from limited data and can be cheaply deployed on constrained hardware. They are user-friendly to train and can attain state-of-the-art performance on a variety of tasks.

This section provides a brief flavor of other ways of approaching two applications, OCR and entity disambiguation. It highlights how problems can be conceptualized differently and underscores some of the advantages of embedding models for academic applications.  

\subsection{Optical Character Recognition}

Section \ref{Ssec:OCR} framed OCR as an image retrieval problem, using a contrastively trained vision model. This diverges from the literature, which mostly models OCR as a sequence-to-sequence (seq2seq) problem. Seq2seq models transform one sequence of data into another sequence. They are frequently used when the input and output data are sequences that may differ in length, such as in machine translation.

Figure \ref{fig:arch} highlights the differences between the \effocr and seq2seq architectures for OCR. First, seq2seq OCR typically requires line-level inputs and does not localize individual characters or words. Instead, it divides text line images or their representations into fixed-size patches. In contrast, \effocr uses modern object detection methods \citep{cai2018cascade, Jocher_YOLOv5_by_Ultralytics_2020} to locate characters or words in the input image.

Second, seq2seq decodes image representations---created by a learned vision model---into text sequentially using a learned language model. Conversely, \effocr employs contrastive training \citep{khosla2020supervised} to learn a meaningful metric space for OCR. Its vision model projects crops of the same character or word close together, regardless of style, while projecting crops of different characters or words to dissimilar embeddings. With \effocr, the vision embeddings alone are sufficient to infer text, by retrieving their nearest neighbor from an index created by embedding a digital font. In contrast, with seq2seq, the vision representations are decoded into texts with a language model, which requires jointly estimating millions of additional parameters.

A drawback of the seq2seq architecture is that it is challenging to extend and customize to novel settings \citep{hedderich-etal-2021-survey}, because training a joint vision-language model requires a vast collection of labeled image-text pairs and significant compute, particularly when state-of-the-art architectures are used. TrOCR—a transformer seq2seq model created by researchers at Microsoft—was trained using 684 million English synthetic text lines and 32 32GB V100 GPUs, a very costly setup that no academic researcher can replicate.

This drawback can be quantified by sample efficiency, which refers to how well a model can perform following exposure to a limited number of training examples. Some architectures learn more efficiently than others. Bi-encoders tend to learn efficiently, which is important for economists because our compute and annotation budgets are severely limited by deep learning standards.

Figure \ref{fig:efficiency}, drawn from \cite{carlson2023}, examines sample efficiency by training various open-source OCR architectures on the same small training sets.
The x-axis plots the percentage of the \effocr training dataset used in training, and the y-axis plots the character error rate.

\input{figures/SampleEfficiency.tex}

On just 99 labeled table cells for Japanese tables and 21 labeled rows from U.S. newspapers in the Chronicling America collection, as well as digital fonts (the 5\% train splits), \effocr's character error rate is around 4\%, showing viable few-shot performance.  
Other architectures, trained on identical data, remain unusable. 
\effocr performs nearly as well using 20\% of the training data as using 70\%, where it continues to outperform all other alternatives. TrOCR learns almost nothing from the amount of data we can expose it to (hence why Microsoft trained it on 684 million text lines). In contrast, \effocr can be trained with a student account in the cloud, or even on a laptop. CRNN is a much lighter weight, older seq2seq architecture, that learns better with limited data but uses an LSTM rather than a transformer, leading to an accuracy hit when fully trained.

Embedding models can also offer a computational advantage over seq2seq models. 
\effocr supports inference parallelization across characters, promoting faster inference, whereas seq2seq requires autoregressive decoding, which is slower. \effocr runs approximately 50 times faster than TrOCR, the only open-source model comparable in accuracy once fully trained. We created the open-source American Stories dataset consisting of over 430 million historical news articles with a \$60,000 budget using \effocr, and could not have realistically generated any substantial additional funds, let alone 50 times more funds.

A sample and compute efficient architecture makes it possible to achieve high-quality transcriptions in diverse settings and on massive-scale document collections, bringing big data to a diversity of economic history applications. In theory, contextual understanding from the full sequence of representations could lead to better OCR. In practice, state-of-the-art transformer seq2seq models are expensive to train and deploy, and are not available for lower-resource languages, with advances mainly in a few languages. By moving away from seq2seq models, significant improvements in sample and computational efficiency can be achieved. In an academic setting, such advantages are particularly relevant, as our applications are highly diverse and our budgets are usually extremely constrained.

\subsection{Entity Disambiguation}

Entity disambiguation---linking entity mentions in unstructured texts to an external knowledgebase like Wikipedia---has led to the development of various architectures. These include a masked language model (LUKE, \cite{yamada2022global}) and a neural translation model (GENRE, \cite{de2020autoregressive})---which uses a sequence-to-sequence architecture to translate mentions into Wikipedia ids---as well as the bi-encoder embedding architecture that treats entity disambiguation as a nearest neighbor retrieval problem \citep{wu2019scalable}. The latter architecture is used by this article's application (Section \ref{Ssec:entity_disambig}). 

The masked language model approach masks out entities and predicts their Wikipedia id using a classifier head. 
While it is the leader on some benchmarks, in practice it has major limitations. 
Language models predict masked tokens through classification.
LUKE is restricted to the top 50K Wikipedia entries, due to computational constraints in calculating the softmax.  Many of the top-50K entries are not people, and many people that appear in historical news or government documents are not amongst the top-50K. 
Additionally, it does not accommodate out-of-knowledge base entities and requires sparse entity priors to initialize the model. In many applications, not all individuals will be in a knowledge base, and the model needs to be able to predict this.

The neural translation model's sequence-to-sequence architecture is slow during inference, taking approximately 60 times longer to run than bi-encoder embedding models. \cite{arora2024contrastive} also show that an embedding model achieves higher accuracy disambiguating historical texts. In short, again a sequence-to-sequence architecture is costly to run at scale and doesn't necessarily offer performance advantages. 

\section{Conclusion} \label{Sec:conclude}

Deep learning provides powerful tools for processing unstructured data. On social science tasks ranging from text classification to record linkage, entity disambiguation, and tracing the spread of reproduced content, deep learning can outperform traditional sparse methods by a wide margin (often by 20 points of F1/accuracy or more) \citep{silcock332noise, dell2023american, arora2023linktransformer, arora2023linking, arora2024contrastive}. 

Deep learning can facilitate novel analyses by providing tools to impute structured information from unstructured data on a massive scale. When working with lightweight, sample-efficient pre-trained models, training and deployment are quite affordable, even for datasets with millions or billions of observations. For some applications, deep learning also offers promise for processing data from low-resource settings, with the potential to make economic research more representative of the diversity of human societies.

Becoming familiar with deep learning methods entails significant startup costs. This article---along with the accompanying open-source packages, tutorials, and knowledge base---aims to significantly reduce entry barriers for economists who would like to use deep learning in their research.

\bibliographystyle{aea}
\bibliography{dl} 

\section{Appendix}

{\renewcommand{\arraystretch}{1.8}%
\centering
\begin{table*}[ht]
\caption{Prompts used to query OpenAI GPT models}
\begin{tabular}{p{0.25\linewidth}p{0.65\linewidth}} 
    \textbf{Topic}&  \textbf{Prompt}\\
    \midrule
    advice&  We would like to classify whether a text is from an advice column. Advice columns answer letters from a reader seeking advice, or give unsolicited advice to readers. Examples include Dear Abby, Ask Ann Landers, Dear Doctor, etc. Yes/No: \\ 
    antitrust&  We would like to classify whether a text is about antitrust action. An article that is about antitrust action covers business practices that stifle competition, or accusations of such practices. It might involve legal action, government regulation, the breaking up of monopolies, or any plans to do so. Yes/No:\\ 
    bible&  We would like to classify whether a text reproduces a short religious blurb, like a quote from the Bible, prayer for the day, spiritual thought for the day, without explanations, opinions, interpretations, or discourses. Long sermons, discourse, or longer texts quoting from the Bible won't count as this `short religious blurb.' This will generally look like `X for the day' - short and crisp with no explanations or opinions at all besides the small blurb. Yes/No:\\ 
    civil rights movement&  We would like to classify whether a text is about the Civil Rights movement. This includes articles referring to organizations and individuals that protested racism against Black Americans, and articles discussing racial discrimination in the government, social and educational inequality of African Americans, police brutality against African Americans, the use of federal power to protect civil rights, or segregation. Articles about protests and riots are on topic if they stemmed from conflicts over civil rights or occurred after the death of Martin Luther King Jr., but race riots and acts of violence involving African Americans should not be on topic if they do not refer to discrimination. Some articles may refer to states rights or express anger about rioting—these articles are on topic only if race or federal protection of civil rights is mentioned. Yes/No:\\ 
    contraception&  We would like to classify whether a text is about contraception. Abortion is not contraception. Yes/No:\\ 
    \bottomrule
    \label{tab:prompts-1}
\end{tabular}
\end{table*}
}

{\renewcommand{\arraystretch}{1.8}%
\centering
\begin{table*}[ht]
\begin{tabular}{p{0.25\linewidth}p{0.7\linewidth}} 
    \textbf{Topic}&  \textbf{Prompt}\\
    \midrule 
     crime&  We would like to classify whether a text covers about crime. This includes reports of crimes and investigations, coverage of court proceedings, law enforcement, and discussions of crime prevention and community safety. Violations of international law are not considered crimes. Nor are actions that may be unethical but are not illegal. Articles about Watergate should not be classified as about crime. Yes/No: \\ 
    horoscope&  We would like to classify whether a text is a newspaper horoscope. Horoscopes are articles that make predictions about people's futures based on astrological signals or signs (sun signs for instance) and are written to entertain and predict future events. They can be of daily frequency, monthly or even yearly. They can also be making astrological predictions for celebrities and known personalities. Yes/No: \\ 
    labor movement&  We would like to classify whether a text covers any American labor movements. A text that is on topic may cover unions striking, advocating for workers’ rights, lobbying the government, or speaking about the experience of working in their industry. It may also cover statements by employers that express anger at union action. Articles that express support for or criticize the labor movement should both count as on topic. Yes/No:\\ 
    obituaries&  We would like to classify whether a text is an obituary. An obituary is defined as an article which mentions a death of a person and provides their Biographical/Family information or Funeral information. Articles which just mention deaths of individuals in passing and do not summarize information about their life or their funeral services are not obituaries. Yes/No:\\
    pesticide& We would like to classify whether an article is about pesticides. Texts are on topic if they mention any chemical attempt to deter pests (insects, other animals, or fungi) that were on a crop or plant. This does not include herbicides, killing insects not on crops, non-chemical bug killing, or anything else about insects. Yes/No:\\
    polio vaccine& We would like to classify whether a text is about the polio vaccine. Yes/No:\\
    \bottomrule
    \label{tab:prompts-2}
\end{tabular}
\end{table*}
}

{\renewcommand{\arraystretch}{1.8}%
\centering
\begin{table*}[ht]
\begin{tabular}{p{0.25\linewidth}p{0.7\linewidth}} 
    \textbf{Topic}&  \textbf{Prompt}\\
    \midrule 
    politics& We would like to classify whether a text is a political news. Yes/No:\\
    protests& We would like to classify whether a text is about protests. Yes/No:\\
    Red Scare& We would like to classify whether a text is about the Red Scare or McCarthyism. Any text that reports concerns about Communist infiltration into the U.S. government or institutions will be on topic. In addition, any accusations of U.S. citizens display communist sympathies will be on topic. References to the Korean War or the expansion of Communism abroad are not on topic, though fears of espionage in the U.S. by foreign Communists agents are on topic. Articles casting doubt on the extent of Communist influences in the U.S. should also be on topic. Yes/No:\\
    schedules& We would like to classify whether a text is a schedule. This includes TV/Game/Movie/Church and other schedules. Schedules list multiple events or programs on the same medium or by the same organization. They are generally lists of time stamps of events followed by no (or a one line) description about them. A church calendar is a schedule, but a summary of the Sunday service is not. Movie listings are a schedule but a special screening of a particular movie is not. Yes/No: \\
    sports& We would like you to classify whether a text is about sports. This does not include fishing, hunting, or flying. Yes/No:\\
    \bottomrule
    \label{tab:prompts-3}
\end{tabular}
\end{table*}
}

{\renewcommand{\arraystretch}{1.8}%
\centering
\begin{table*}[ht]
\begin{tabular}{p{0.25\linewidth}p{0.7\linewidth}} 
    \textbf{Topic}&  \textbf{Prompt}\\
    \midrule 
    Vietnam War& We would like you to classify whether a text pertains to (any aspect of) the Vietnam War. Yes/No:\\
    weather& We would like to classify texts that contain measurements about the weather. These include the weather forecast as well as texts summarizing what the weather was in the recent past (e.g., temperature, precipitation, visibility). This also includes articles about weather records, like the coldest/hottest days in a decade/year or coldest/hottest places. If the article contains weather measurements but also talks about the consequences of the weather in passing (e.g., damage, events cancelled), we will call it about weather measurements. However, if it just talks about the consequences of the weather (e.g., closures, accidents, damage) without giving the weather measurements, this does not count. As a rule of thumb, if there is precise weather measurement somewhere in the article- call it a weather measurement article. Yes/No:\\
    World War I& We would like to classify whether a text is about World War I. Yes/No:\\
    \bottomrule
    \label{tab:prompts-4}
\end{tabular}
\end{table*}
}
\end{document}

%% file: figures/flowchart.tex
\begin{figure*}[!ht]
    \centering
 \includegraphics[width=.8\linewidth]{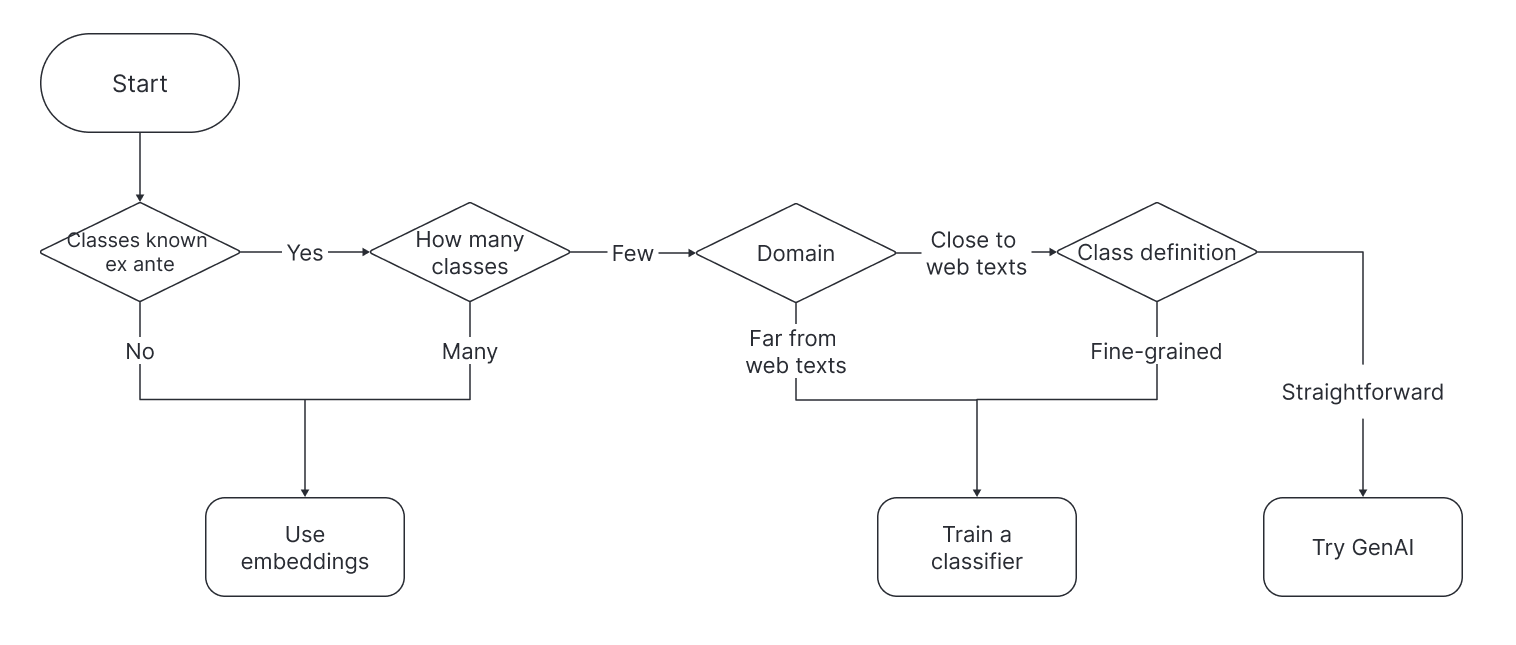}
 \caption{Flowchart for approaching classification.}
    \label{flowchart}
   \end{figure*}

%% file: tables/applications.tex
\begin{table}[!ht]
\caption{Applications}
    \resizebox{\linewidth}{!}{
       \begin{tabular}{llll}
        \toprule
        \textbf{Problem} & \textbf{Modality} & \textbf{Application(s)} & Section \\
        \midrule
            \multicolumn{3}{l}{\textbf{\textit{Classifiers and GenAI}}} \\
        Sequence   & Text & Classify news & \ref{ssTopicClassification}  \\
        \ \ classification & & article topics \\  \\ 
        Token  & Text & Tag people, & \ref{Ssec:NER} \\ 
        \ \ classification & & locations, orgs \\ \\
        Paired text  & Text & Text b & \ref{Ssec:cross-encoder} \\ 
        \ \ classification & & entails a? \\ \\
        \multicolumn{2}{l}{\textbf{\textit{Embedding Models}}} \\
        Link   & Text, & Link & \ref{Ssec:LinkLLM} \\ 
        \ \ structured & Images & firms, products, \\ 
        \ \ data & & locations \\ \\
         Link   & Text &  Link people & \ref{Ssec:entity_disambig} \\ 
        \ \ unstructured & &  mentions to \\
        \ \ data & & Wikipedia \\ \\
          Classification   &  Text & Track content & \ref{Ssec:wireclustering}  \\  
       \ \ w/  unknown  & Images & dissemination; \\
      \ \    categories & & data exploration \\ \\
      Retrieval & Images & Optical character & \ref{Ssec:OCR} \\
      & & recognition \\ \\
        \textbf{\textit{Regression}} \\
         Object & Images & Detect document & \ref{Sec:regression}  \\ 
        \ \ detection &  & layouts \\
        \bottomrule
    \end{tabular}}
    \begin{tablenotes}
    Applications covered in this review.
    \end{tablenotes}
    \label{tab:apps}
\end{table}

%% file: figures/bow.tex
\begin{figure*}[ht]
    \centering
    \includegraphics[width=.9\linewidth]{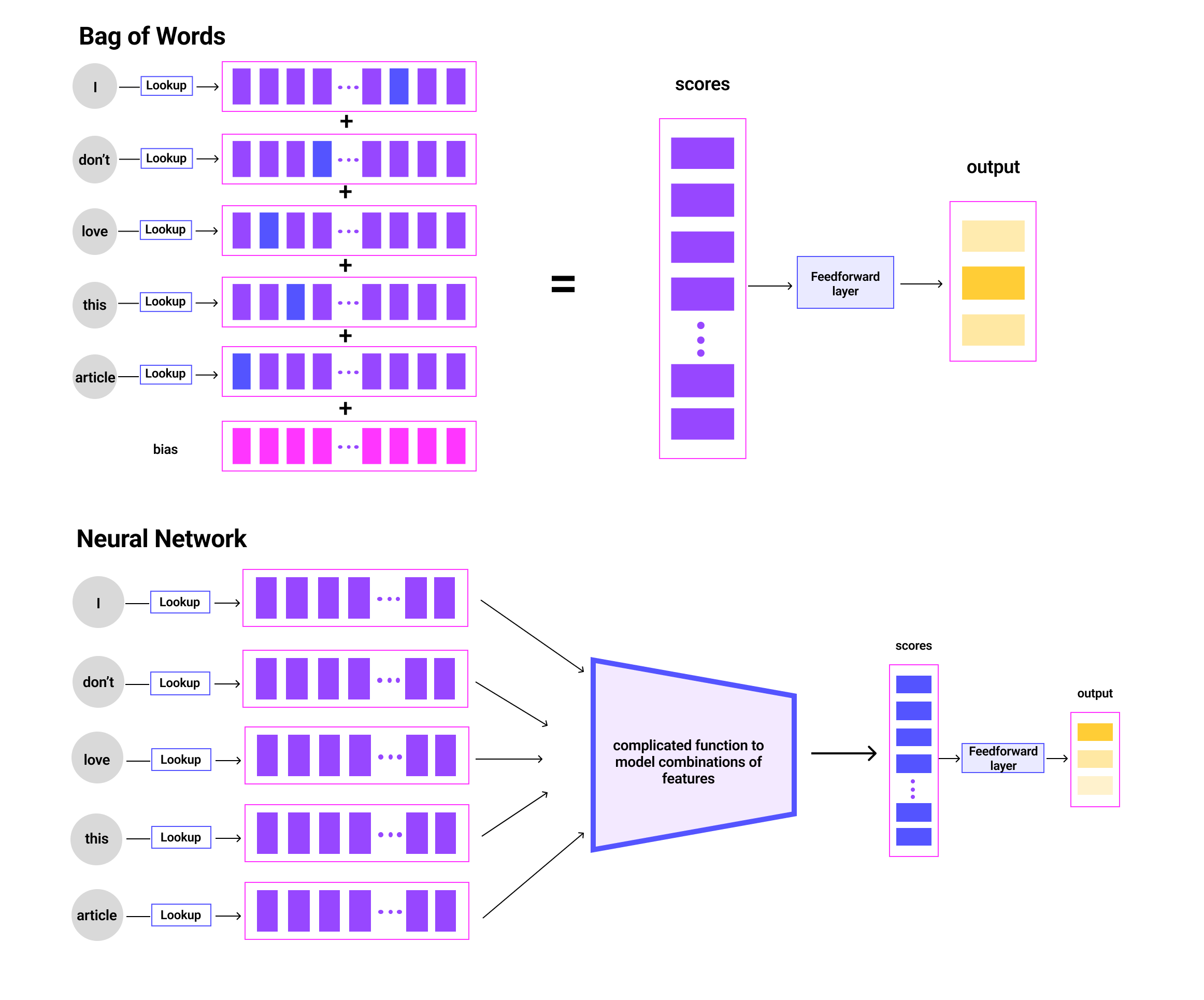}
    \caption{Classifying text.} 
    \label{fig:bow}
    \vspace{-4mm}
  \end{figure*}

%% file: figures/BERT.tex
\begin{figure*}[htb]
\centering
\includegraphics[width=.7\linewidth]{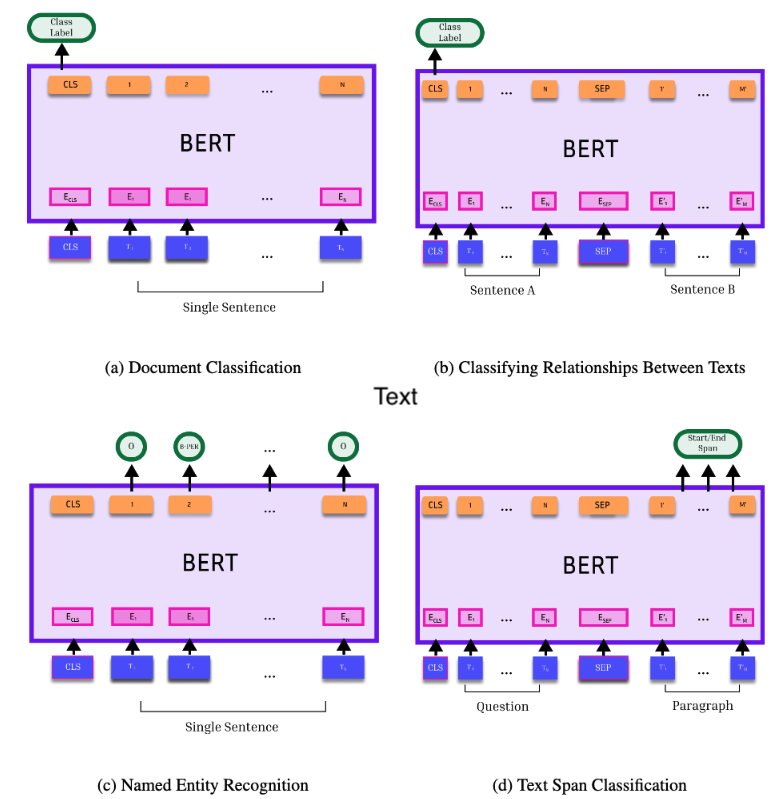}

\caption{Tasks performed with a transformer encoder language model.} \label{fig:bert}
\end{figure*}

%% file: figures/NER.tex
\begin{figure*}[ht]
    \centering
    \includegraphics[width=.95\linewidth]{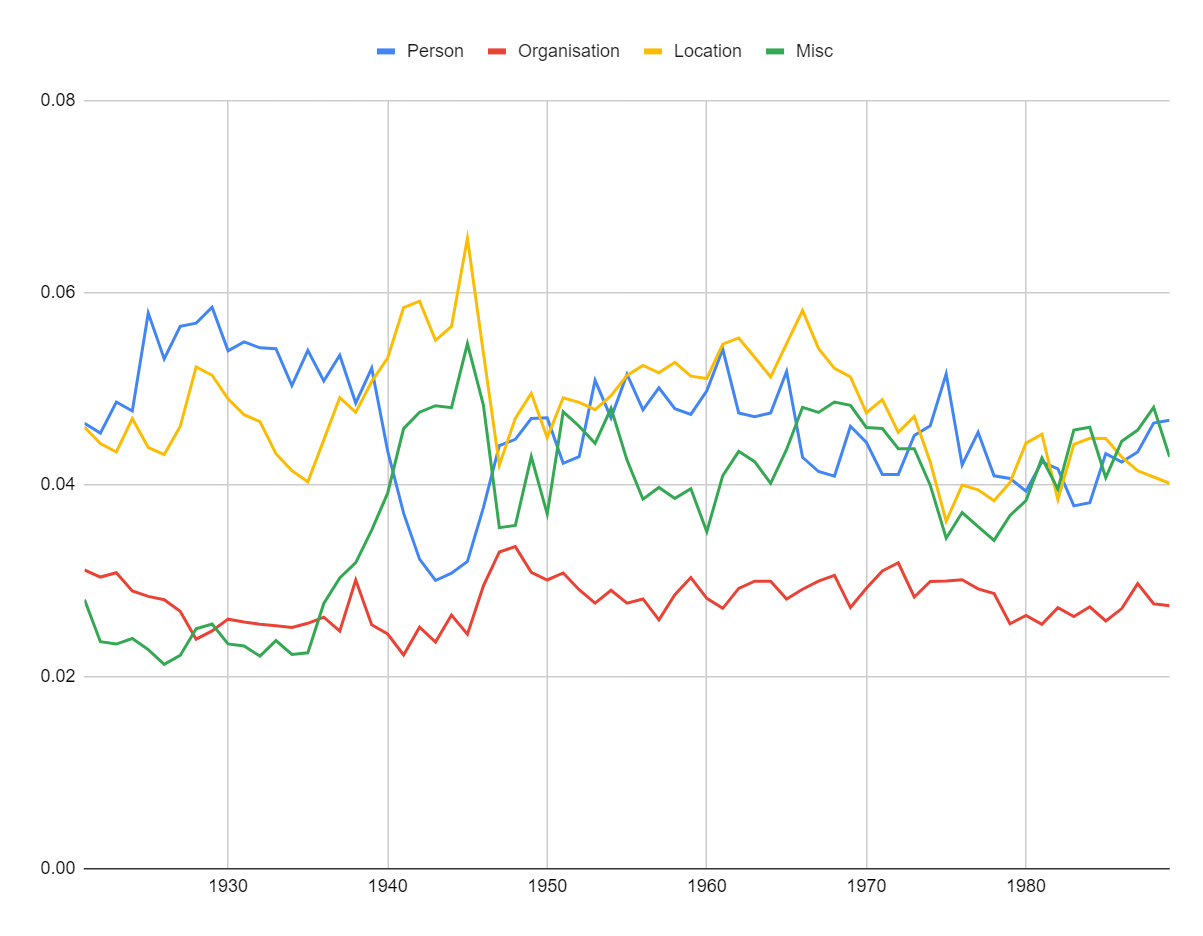}
    \caption{Shares of entity types in newswire articles.} 
    \label{fig:ner}
    \vspace{-4mm}
  \end{figure*}

%% file: figures/cross_encoder.tex
\begin{figure}[ht]
    \centering
    \includegraphics[width=0.5\linewidth]{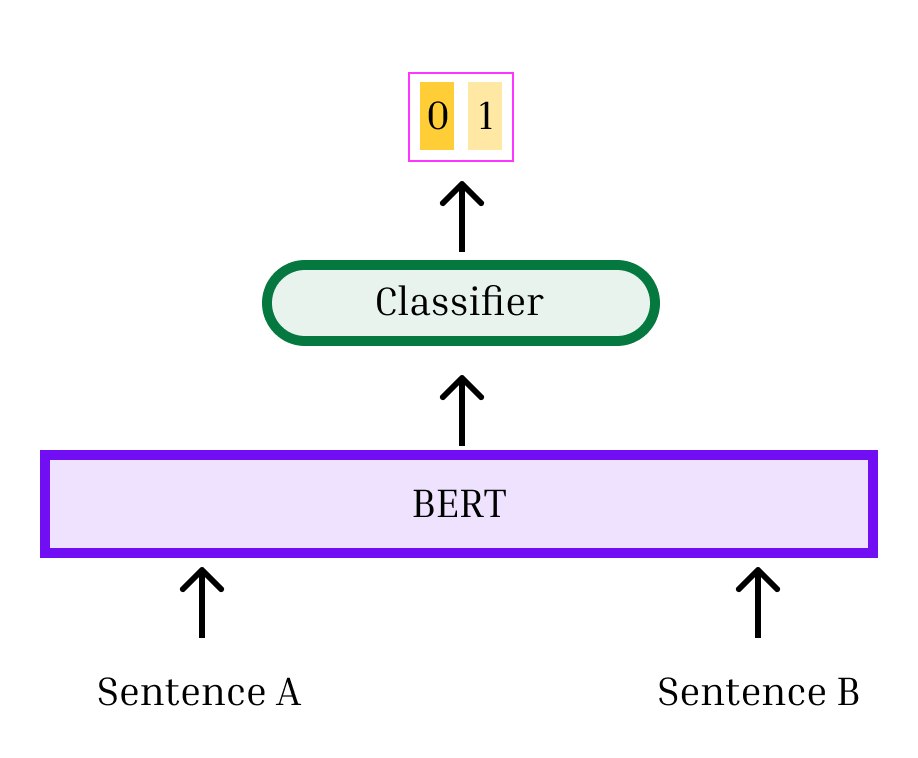}\\
    Cross-encoder\\
    \includegraphics[width=\linewidth]{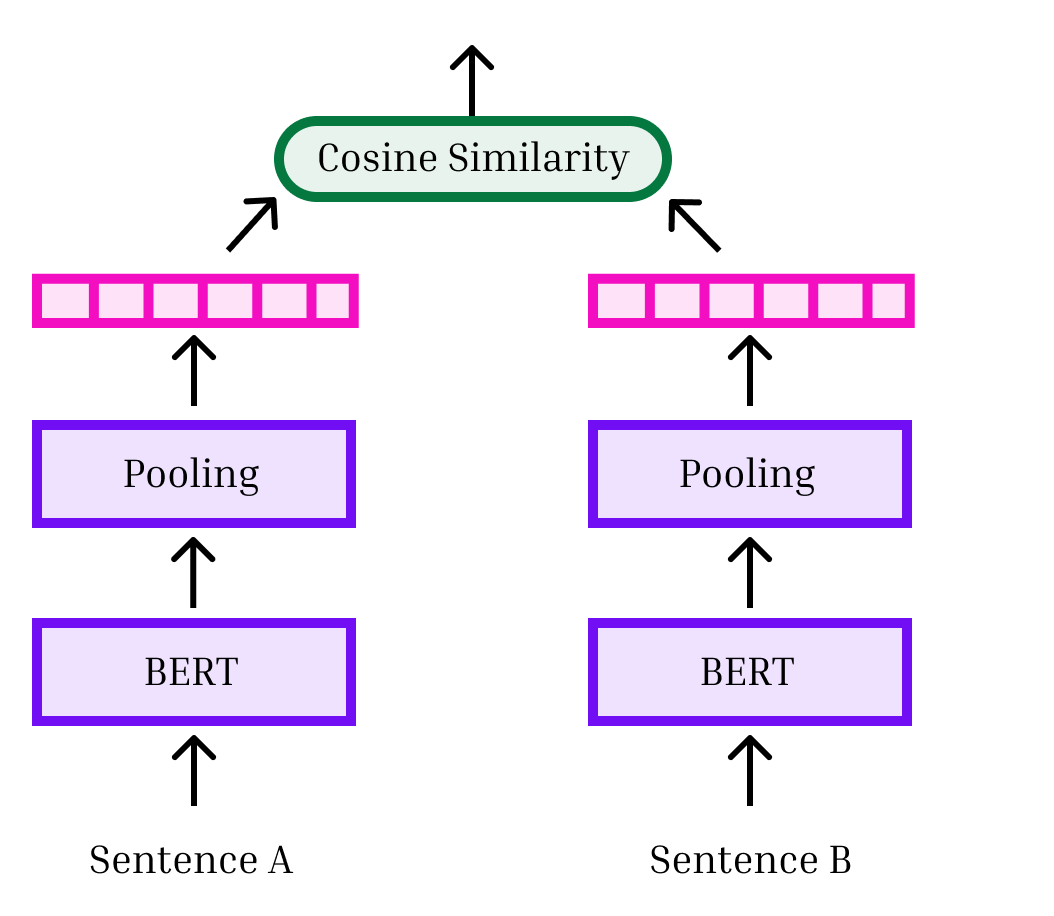}\\
    Bi-encoder
    \caption{Architectures to compare texts.} 
    \label{fig:cross_bi}
    \vspace{-4mm}
  \end{figure}

%% file: figures/ComparativeAgendas.tex
\begin{figure*}[!htb]
\centering
\includegraphics[width=\linewidth]{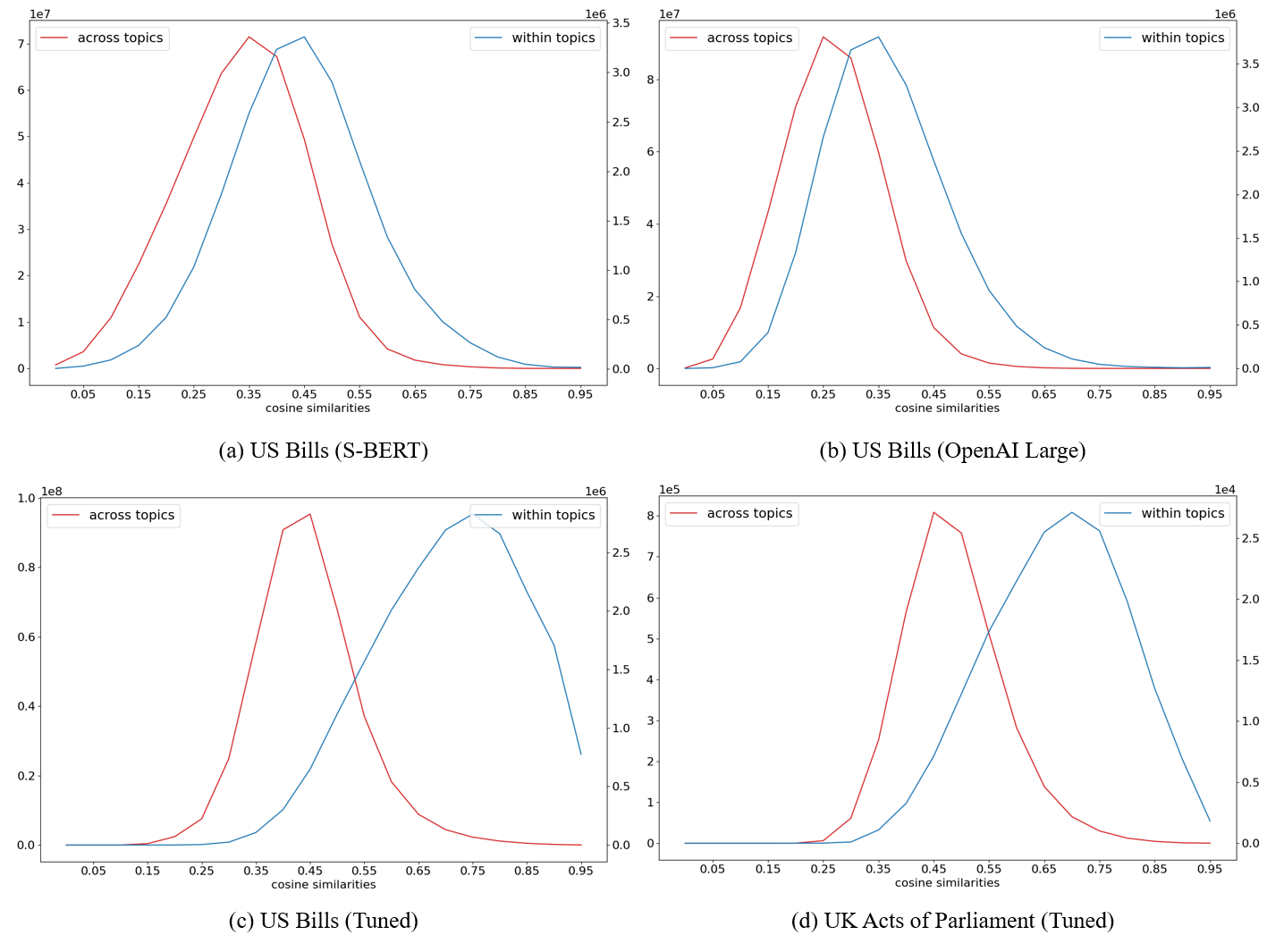}

\caption{Embedding similarities within and across topics.} \label{fig:comparative_agendas}
\end{figure*}

%% file: figures/char_near.tex
\begin{figure}[ht]
    \centering
    \includegraphics[width=0.95\linewidth]{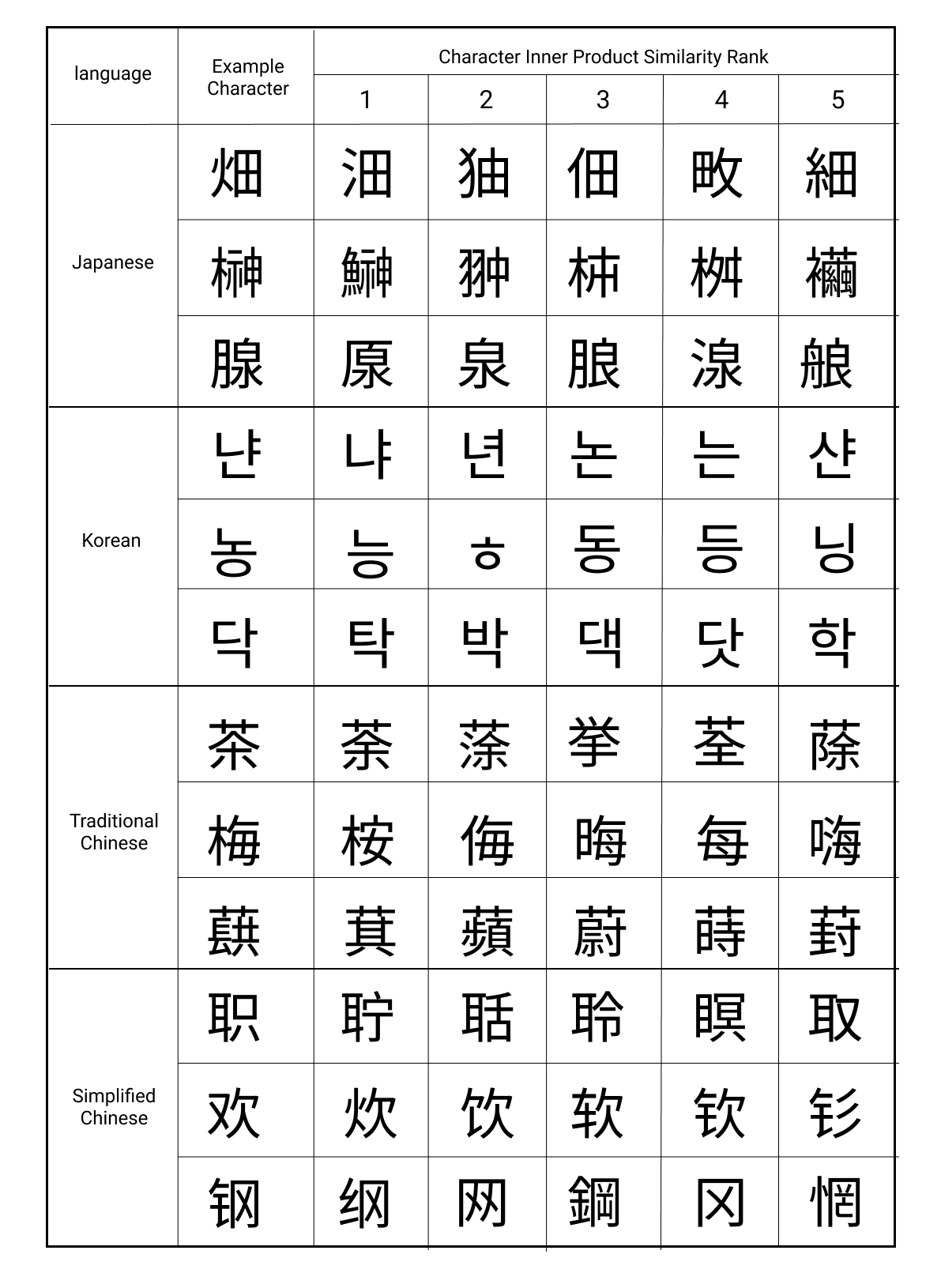}\hfill
   \caption{Character similarity, as measured by vision transformers.} 
    \label{fig:char_near}
    \vspace{-4mm}
  \end{figure}

%% file: figures/ca_layouts.tex
\begin{figure*}[htb]
    \centering
\includegraphics[width=\linewidth]{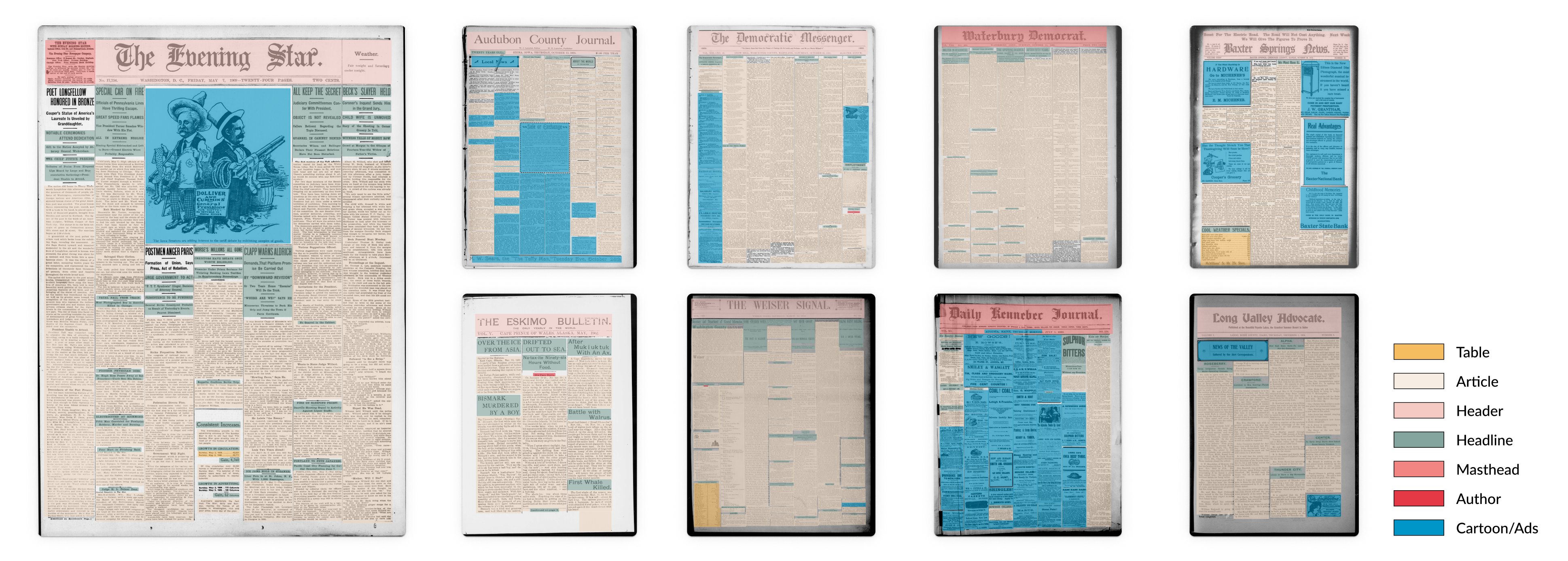}
\caption{Layouts from historical newspapers.}
\label{fig:ca_layouts}
  \end{figure*}

%% file: figures/gcv.tex
\begin{figure}[hbt]
    \centering
    \includegraphics[width=\linewidth]{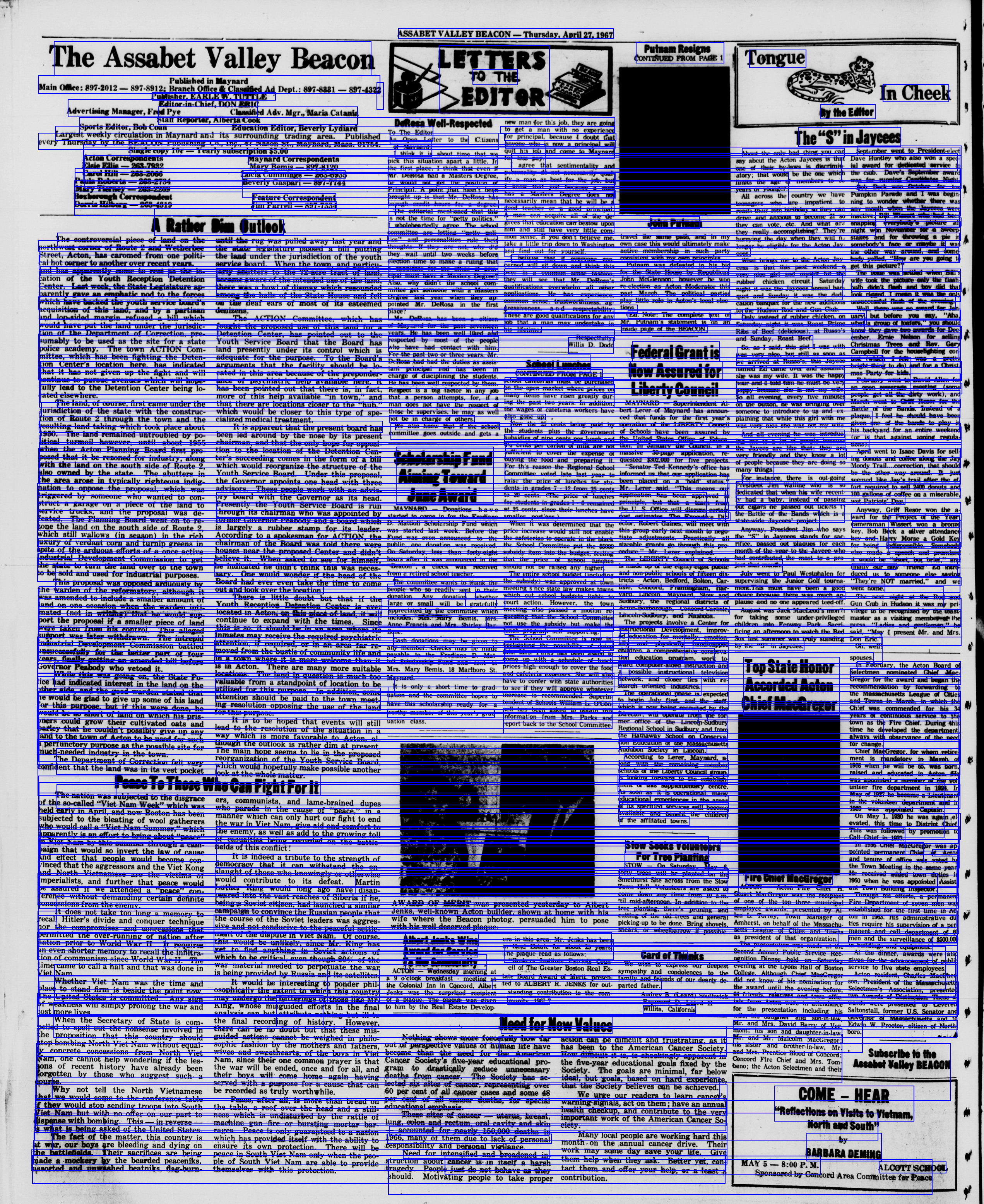}
    \caption{Layouts as detected by Google Cloud Vision.}
    \label{gcv}
  \end{figure}

%% file: figures/tk_layouts.tex
\begin{figure}[ht]
    \centering
    \includegraphics[width=\linewidth]{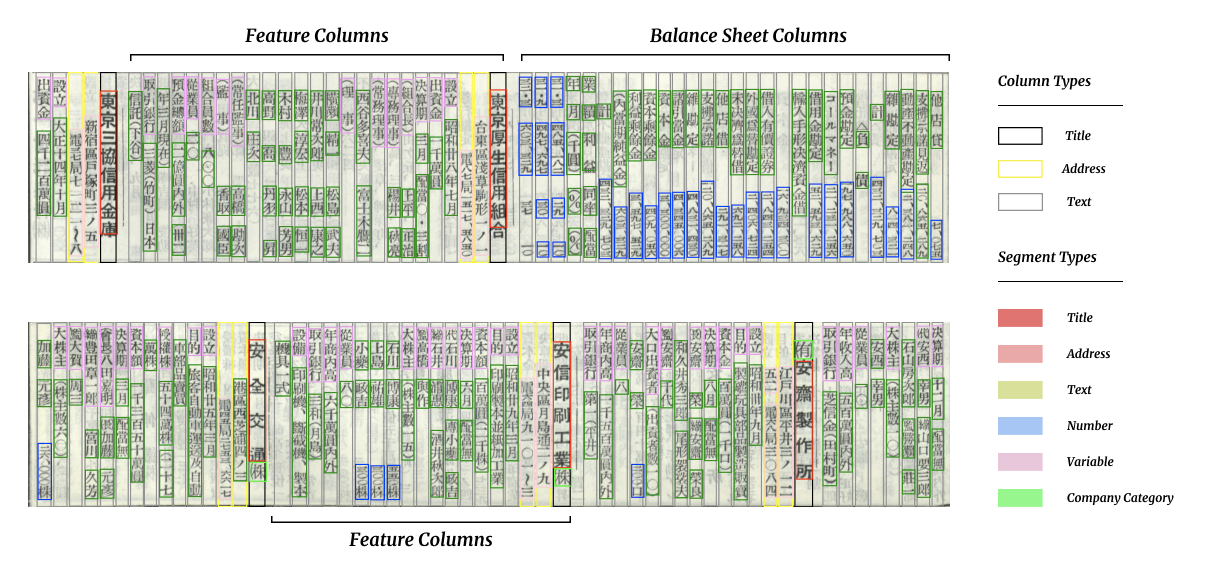}
    \caption{Document layouts.} 
    \vspace{-2mm}
    \label{fig:tk}
  \end{figure}

%% file: figures/EffOCR_arch.tex
\begin{figure*}[ht]
    \centering
    \includegraphics[width=.95\linewidth]{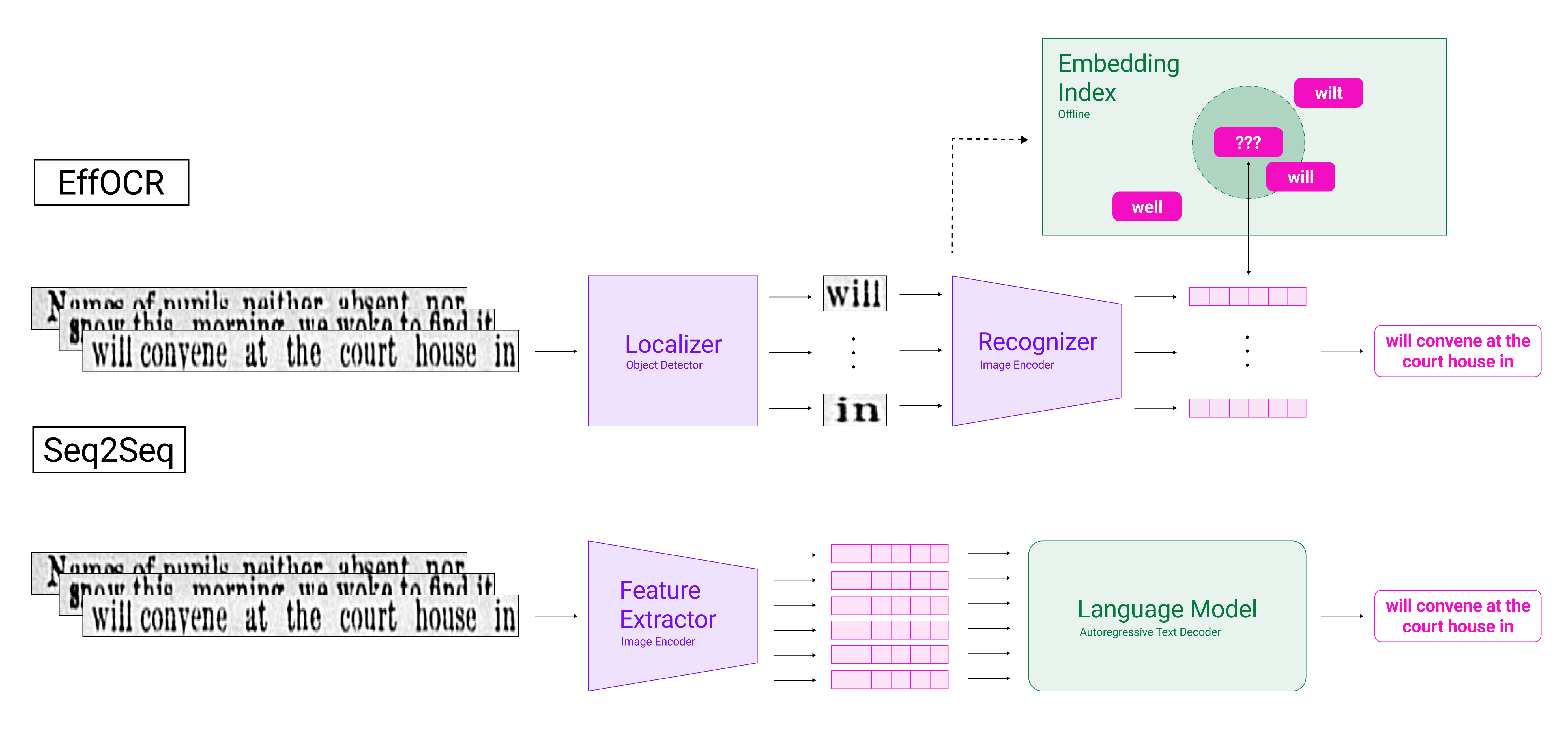}
    \caption{EffOCR and Seq2Seq Model Architectures.} 
    \label{fig:arch}
    \vspace{-4mm}
  \end{figure*}

%% file: figures/SampleEfficiency.tex
\begin{figure}[ht]
    \centering
    \includegraphics[width=\linewidth]{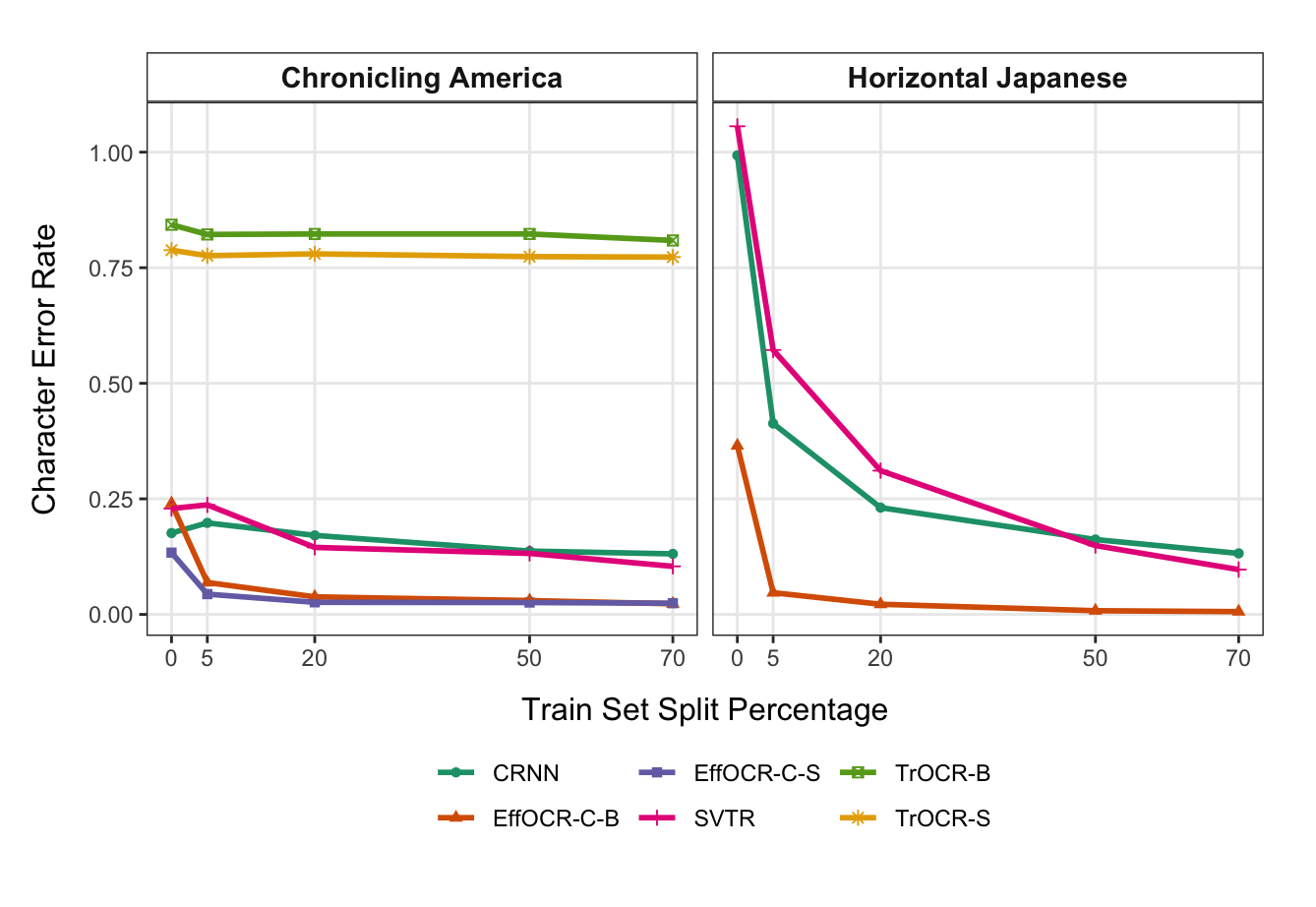}
    \caption{Sample Efficiency.} 
    \label{fig:efficiency}
    \vspace{-4mm}
  \end{figure}